\documentclass[aps,prb,twocolumn,superscriptaddress,10pt]{revtex4-2}
\pdfoutput=1 
\usepackage[utf8]{inputenc}
\usepackage[english]{babel}
\usepackage[T1]{fontenc}
\usepackage[pdftex]{graphicx}
\usepackage{amsmath}
\usepackage{amsthm}
\usepackage{amssymb}
\usepackage{braket}
\usepackage{xcolor}
\usepackage{dcolumn}
\usepackage{bbm}
\usepackage{bbold}
\usepackage[normalem]{ulem}
\usepackage[colorlinks=true,linkcolor=blue,citecolor=blue,urlcolor=blue,unicode]{hyperref}
\usepackage{footmisc}
\usepackage{physics} 

\newcommand{\Luc}{L_{\text{uc}}}

\newcommand{\red}[1]{{\color{red} #1}}

\renewcommand{\red}[1]{{#1}}
\renewcommand{\sout}[1]{}

\begin{document}
\title{Robustness and eventual slow decay of bound states of interacting microwave photons in the Google Quantum AI experiment}
\author{Federica Maria Surace}
\affiliation{Department of Physics and Institute for Quantum Information and Matter,
California Institute of Technology, Pasadena, California 91125, USA}
\author{Olexei Motrunich}
\affiliation{Department of Physics and Institute for Quantum Information and Matter,
California Institute of Technology, Pasadena, California 91125, USA}

\begin{abstract}
Integrable models are characterized by the existence of stable excitations that can propagate indefinitely without decaying.
This includes multi-magnon bound states in the celebrated XXZ spin chain model and its integrable Floquet counterpart.
A recent Google Quantum AI experiment [A.~Morvan {\it et al.}, Nature {\bf 612}, 240 (2022)] realizing the Floquet model demonstrated the persistence of such collective excitations even when the integrability is broken: this observation is at odds with the expectation of ergodic dynamics in generic non-integrable systems.
We here study the spectrum of the model realized in the experiment using exact diagonalization and physical arguments.
We find that isolated bands corresponding to the descendants of the exact bound states of the integrable model are clearly observable in the spectrum for a large range of system sizes. However, our numerical analysis of the localization properties of the eigenstates suggests that the bound states become unstable in the thermodynamic limit. A perturbative estimate of the decay rate agrees with the prediction of an eventual instability for large system sizes.
\end{abstract}

\maketitle

\section{Introduction}

In recent years, quantum simulators have become a powerful tool to investigate the non-equilibrium dynamics of quantum many-body systems.
Experiments based on many different platforms have now the capability to prepare a quantum state with good fidelity and evolve it almost unitarily, preserving its coherence for sufficiently long times to observe its interacting dynamics~\cite{cirac2012goals,Georgescu2014,Preskill2018quantumcomputingin,bluvstein2021controlling,tan2021domain,daley2022practical}.
These advances sparked a considerable interest in the theoretical investigation of the approach to thermal equilibrium in isolated quantum many-body systems~\cite{rigol2008thermalization,PolkovnikovColloquium,eisert2015quantum, DAlessio2016, gogolin2016equilibration, Mori_2018,AbaninColloquium}.
At the same time, the available quantum simulators inspired the search for ``unusual'' behaviors, where a system does not reach thermal equilibrium.
A paradigmatic example is the phenomenon known as quantum many-body scarring, which was first discovered in an experiment with Rydberg atom arrays~\cite{Bernien2017}: it was found that certain quantum states exhibit long-lived non-ergodic dynamics, while other states instead show fast thermalization.
Following the initial discovery, numerous theoretical studies have been conducted to elucidate the conditions under which quantum many-body scars occur and to develop a theoretical framework that can explain and predict such phenomena~\cite{Turner2017,Moudgalya2018a,Shiraishi2017,Serbyn2020review,Papic2021review,Moudgalya2021review,Chandran2022review}.
Other examples of exotic dynamical phenomena that have been observed in quantum simulators include, for instance, Hilbert space fragmentation \cite{Sala2020,Khemani2020,Kohlert2023}, dynamical phase transitions \cite{Heyl2013,flaschner2016observation,zhang2017dpt}, time crystals \cite{Khemani2016,Else2016,Yao2017,zhang2017observation,choi2017observation,khemani2019brief,mi2022time}, and noise-resilient edge modes \cite{fendley2016strong,mori2016rigorous,abanin2017rigorous,Else2017,Else2017b,Kemp_2017, Mi2022edge}.

\begin{figure}[t]
    \centering
    \includegraphics[width=\linewidth]{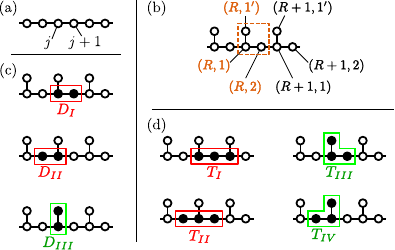}
    \caption{(a) 1d chain hosting the integrable Hamiltonian and Floquet systems. 
    (b) 1d chain with extra sites forming a comb in the GQAI experiment, with the comb teeth connected to every other site of the original chain;
a unit cell and labelling of sites in each cell is shown. 
    (c) Schematic of distinct two-particle bound states, which become dimers in the very strong binding limit, in a unit cell on the comb; these give rise to three bands in the momentum space. 
    (d) Distinct three-particle bound states, schematized as trimers, in a unit cell on the comb, producing four bands in the momentum space.}
    \label{fig:lattice_and_boundstates}
\end{figure}

Another puzzle for our understanding of quantum many-body dynamics was recently observed in a Google Quantum AI (GQAI) experiment \cite{morvan2022formation,Prosen2022} based on superconducting circuits hosting microwave photons.
The dynamics of the photons, which can hop between neighboring sites of a chain  shown in Fig.~\ref{fig:lattice_and_boundstates}(a)  and interact with each other, is described by an integrable quantum circuit.
The experiment showed the presence of bound states of up to 5 photons: these are exact eigenstates of the model predicted by Bethe ansatz~\cite{Aleiner2021}, where the photons are nearly adjacent and form a single collective excitation.
The bound states observed in the study are enclosed in the continuous spectrum of multi-particle states, i.e., eigenstates of (mostly) distant photons that can scatter off each other.
Even in the absence of a gap, the conservation laws of the integrable model protect the bound states from mixing with the underlying continuum.
The work~\cite{morvan2022formation} demonstrated a remarkable agreement of the experimental results with the analytical solution.

The integrable circuit was then perturbed by coupling every other site of the chain with additional sites as shown in Fig.~\ref{fig:lattice_and_boundstates}(b), thus breaking the integrability of the model. 
In the absence of any conservation laws that protect them, the exact bound states of the unperturbed model are expected to quickly decay in the continuum of multi-particle states.
Nevertheless, the experiment showed the persistence of the bound states after many cycles of the circuits, even for fairly large values of the perturbation.

The origin of this robustness cannot be easily attributed to any known mechanism for slow relaxation.
Thus, we do not expect usual quantum many-body scarring mechanisms to be operative in our systems with $O(1)$ integrability breaking, and there is no evidence of even more dramatic Hilbert space fragmentation phenomena.
Among other possibilities, the rigorous theory of prethermalization~\cite{abanin2017rigorous} applies only for perturbing much more special initial Hamiltonians and does not apply for perturbing generic integrable systems like in our case  (except in the regime with dominant on-site terms).
One could contemplate a recently discussed mechanism of weak integrability breaking of integrable models, where special perturbations effectively break the integrability only at higher order~\cite{Kurlov2022,surace2023weak,Orlov2023}.
However, our experience with this mechanism suggests that it would require more complicated special perturbations than in the experiment (and it would approximately preserve features in the entire spectrum and not just some specific states).
While we cannot rule out such interesting and unusual explanations, we are led to consider perhaps simpler mechanisms having to do with particular quantitative properties of the specific system.
Thus, an early example in Ref.~\cite{Banuls2011} of unusual thermalization in a non-integrable model for some initial states was explained in Ref.~\cite{Lin2017} by proximity of these states to the boundaries (ground or ceiling states) of the many-body spectrum; that is, the dynamics was governed by being in a ``special corner'' of the many-body Hilbert space away from ETH-like physics expected in the middle of the many-body spectrum.

While for Floquet systems there is no notion of ground or ceiling states and hence no such proximity can be invoked,
 we note that the circuit and setup that models the experiment in \cite{morvan2022formation}  is not a truly many-body system.
The puzzling robust bound states were observed for a system with a fixed number of photons ($N=3$), and the Hilbert space dimension scales only polynomially with the system size, not exponentially.
It is reasonable to expect that in such few-body models the decay is not as fast as for typical eigenstates in the middle of a truly many-body spectrum.
Therefore, in order to understand and characterize the robustness of the bound states, it is important to perform a quantitative study for increasing system sizes, and to contrast the persistence of the bound states against the scaling expected for similar few-body models.

In this work, we examine the spectral properties of the perturbed model in this few-body regime using exact diagonalization and physical arguments.
We compare the spectrum of the circuit with the one of the  related Hamiltonian system under corresponding continuous time dynamics.
First, we find that in the closely related Hamiltonian case the $N=3$ bound states are in fact protected by a gap and persist in the thermodynamic limit.
This is true also for $O(1)$ integrability breaking perturbations for (a range of) serendipitous choices of parameters, including the Hamiltonian parameters reasonably related to the ones in the experiment.
On the other hand, in the case of the Floquet circuit the bound states are ``folded'' and overlap with the multi-particle continuum in the corresponding Floquet spectrum, but for the values of parameters studied in the experiment and $N=3$ this folding affects only the boundaries of the Hamiltonian spectrum.
Because of this and for other quantitative reasons, despite the folding, we clearly detect isolated bands in the Floquet spectrum that correspond to the unperturbed bound states, even for large values of the perturbation.
Our numerical analysis suggests, however, that these bound states eventually become unstable in the thermodynamic limit.
To argue this, we study the inverse participation ratio in the center of mass frame, whose scaling with system size can distinguish a bound state from a ``scattering'' (i.e., unbound) state.
We show that the inverse participation ratio of the bound states decreases with the system size, indicating a decay in the thermodynamic limit.
However, the decay is very slow: we attribute the slowness to the impossibility in our parameter regime for a 3-photon bound state to resonantly decay into a scattering state of a 2-photon bound state and a single photon; the 3-photon bound state can only decay into a scattering state of $3$ isolated photons, a process with a very small (but non-zero) matrix element. 
Our conclusions are further corroborated by a perturbative analysis of the matrix elements and density of states, which shows a small but finite decay rate in the thermodynamic limit.

The paper is structured as follows.
In Sec.~\ref{sec:model} we introduce the integrable models (Hamiltonian and Floquet circuit) and the non-integrable perturbation.
In Sec.~\ref{sec:spectrum} we analyse the spectra of the Hamiltonian and the Floquet circuit in the sectors with $N=1$, $2$, and $3$ photons, for different values of the perturbation strength.
In Sec.~\ref{sec:instability} we study the properties of the 3-particle bound states in the Floquet circuit for increasing system sizes, and show their eventual instability in the thermodynamic limit.
In Sec.~\ref{sec:perturbative} we study the matrix elements and density of states in the unperturbed model, in order to explain the instability and derive a perturbative estimate of the decay rate.
We conclude in Sec.~\ref{sec:conclusions} with suggestions for more experimental tests and broader prospects.

\section{Hamiltonian and Floquet models}
\label{sec:model}
We will focus on a qubit model that evolves under (i) a static Hamiltonian or (ii) a Floquet circuit.
In both cases, the dynamics we consider conserves the total number of  qubits in the state ``1'' (we will refer to them as ``particles'') and can be expressed in terms of the two-qubit operator $h_{j,k}(w,u)$ acting on neighbouring sites:
\begin{equation}
h_{j,k}(w,u) = - w (\ketbra{10}{01}_{j,k} + \mathrm{H.c.}) - u \ketbra{11}{11}_{j,k} ~, 
\end{equation}
where $w$ is the hopping amplitude between sites $j$ and $k$ and $u$ is an interaction (attraction energy for $u>0$) when particles occupy these sites.

The Hamiltonian and the Floquet circuit that we will consider are perturbations of an integrable Hamiltonian/Floquet model, respectively.
We will first introduce the integrable models (Sec.~\ref{sec:Hint} and \ref{sec:Fint}), and then consider the non-integrable perturbation (Sec.~\ref{sec:HFnonint}).

\subsection{Hamiltonian: integrable case}
\label{sec:Hint}
The integrable chain Hamiltonian with periodic boundary conditions (assumed throughout) is simply
\begin{equation}
\label{eq:Hint}
H_0 = \sum_j h_{j,j+1}(w,u) ~.
\end{equation}
The Hamiltonian $H_0$ can be equivalently written in terms of spin $1/2$ operators, defined as 
$S^z = (\ketbra{1} - \ketbra{0})/2$,
$S^x=(\ketbra{1}{0} + \ketbra{0}{1})/2$, 
$S^y=(-i\ketbra{1}{0} + i\ketbra{0}{1})/2$:
\begin{align}
H_0&= -\sum_j \left[ 2w (S_j^x S_{j+1}^x + S_j^y S_{j+1}^y) + u S_j^z S_{j+1}^z \right] \nonumber\\
&\quad\, +  \sum_j u (S_j^z - 1/4) ~, 
\end{align}
which is the XXZ model in the uniform magnetic field.

\subsection{Floquet circuit: integrable case}
\label{sec:Fint}
An integrable circuit, also known as the Floquet XXZ model, can be defined using the following two qubit gates:
\begin{equation}
\exp[-i h_{j,k}(w,u) t] = \mathrm{fSim}(\theta = w t, \phi = u t)_{j,k} ~,
\end{equation}
where $\mathrm{fSim}$ is the two-site unitary gate using the same notation as in the GQAI experiment~\cite{morvan2022formation}.
The gates are applied to even-odd and odd-even pairs of sites in a brickwall pattern. 
The unitary operator that describes the evolution over a single cycle is defined as
\begin{align}
\label{eq:floq0}
& {\mathcal U}_0(\theta, \phi) = {\mathcal U}_{\mathrm{even}}(\theta, \phi)\; {\mathcal U}_{\mathrm{odd}}(\theta, \phi) ~, \\
& {\mathcal U}_{\mathrm{even}}(\theta, \phi) = \prod_j \mathrm{fSim}(\theta, \phi)_{2j,2j+1} ~, \\
& {\mathcal U}_{\mathrm{odd}}(\theta, \phi) = \prod_j \mathrm{fSim}(\theta , \phi )_{2j-1,2j} ~.
\end{align}
 
The integrability of the model was first demonstrated in \cite{Ljubotina2019}, and a Bethe ansatz solution was obtained in \cite{Aleiner2021}. 
The bound states of the model, that were analytically studied in \cite{Aleiner2021}, were then detected in the GQAI experiment \cite{morvan2022formation}.

\begin{figure*}[t]
\centering
\includegraphics[width=\linewidth]{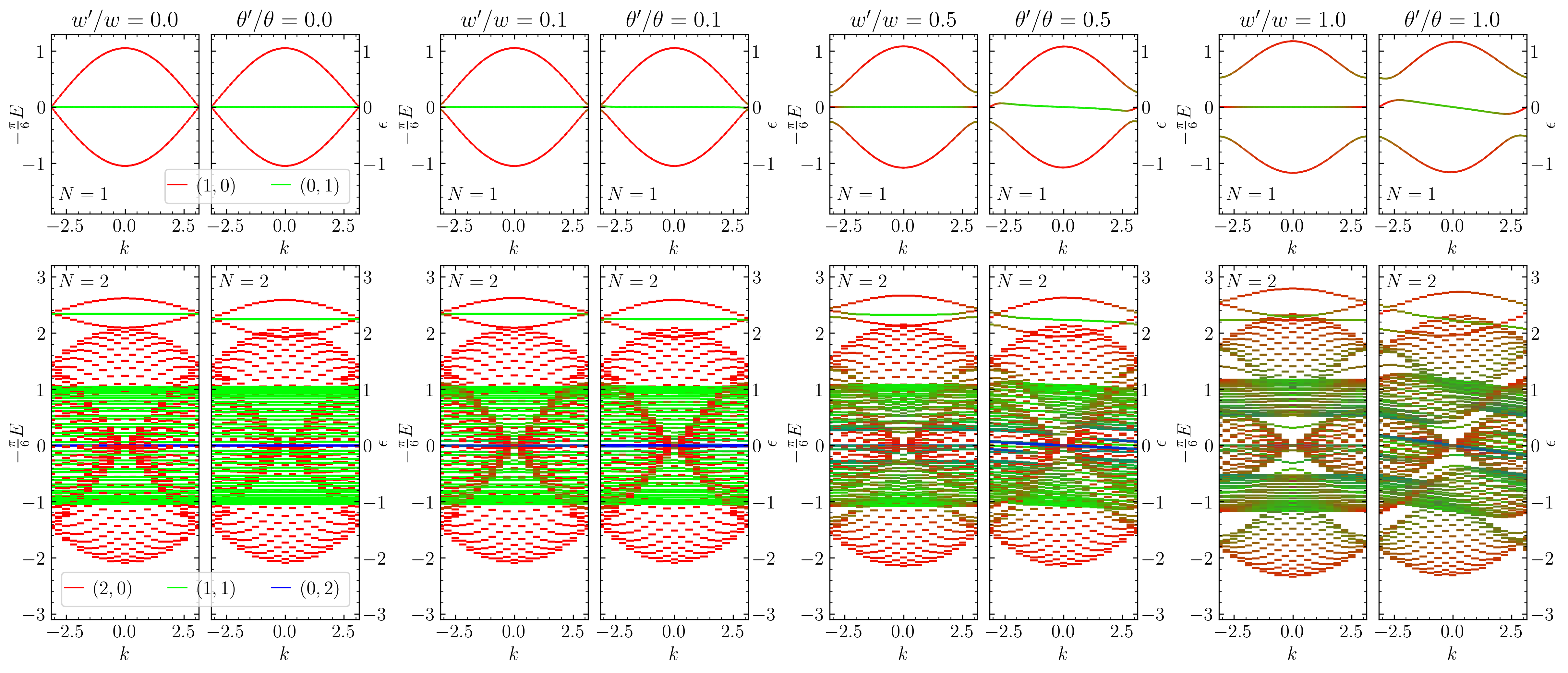}
\caption{Spectrum of the Hamiltonian model (left side in each two-panel group) and Floquet circuit (right side) in the one-particle ($N=1$, top panels) and two-particle ($N=2$, bottom panels) sectors.
The spectra are shown for comparable values of parameters ($w=1$,  $u=u'=4$, $\theta=\pi/6$, $\phi=\phi'=2\pi/3$), and for increasing values of  $w'/w$ and $\theta'/\theta$ taken to be the same in each two-panel group.
The Hamiltonian energies $E$ are multiplied by an appropriate time evolution factor of $(-\pi/6)$ for an approximate match to the Floquet circuit step.
These still fit within a $2\pi$ period and hence can be compared directly with the Floquet quasienergies $\epsilon$ without requiring any folding. \red{For the unperturbed models with $w'/w=0$ and $\theta'/\theta=0$, different colors represent states belonging to different $(N_{1\cup 2}, N_{1'})$ sectors, as shown in the legend. For the perturbed models, $(N_{1\cup 2}, N_{1'})$ are not good quantum numbers, and the eigenstates can have non-zero components in all the sectors:}
The color mixing (red/green for $N=1$ and red/green/blue for $N=2$) represents the squares of the norms of the projections of each eigenstate into the sectors \red{[i.e., RGB color $(255\cdot P(1,0), 255\cdot P(0,1), 0)$ for $N=1$ and $(255\cdot P(2,0), 255\cdot P(1,1), 255 \cdot P(0,2))$ for $N=2$, where $P{(N_{1\cup 2}, N_{1'})}$ is the squared norm of the projection of the eigenstate in the $(N_{1\cup 2}, N_{1'})$ sector]}\red{\sout{ (conserved for $w'/w=\theta'/\theta=0$) with $(N_{1\cup 2}, N_{1'}) = (1,0), (0,1)$ for $N=1$ and $(N_{1\cup 2}, N_{1'}) = (2,0), (1,1), (0,2)$ for $N=2$ respectively}}.
The system size is $\Luc=120$ unit cells for $N=1$, and $\Luc=20$ unit cells for $N=2$.}
\label{fig:spectra12}
\end{figure*}

\subsection{Non-integrable perturbation}
\label{sec:HFnonint}
In Ref.~\cite{morvan2022formation} the integrable model was perturbed by adding sites as in Fig.~\ref{fig:lattice_and_boundstates}(b). 
The additional sites are connected in the shape of ``comb teeth'' to every other site of the original chain.
To label the sites in the new geometry, we use a composite index $(R,\alpha)$ where $R$ labels the unit cell and $\alpha \in \{1, 2, 1' \}$ labels the three sites of each unit cell. 
In this new geometry, we can write perturbed models for both the Hamiltonian and the Floquet case.

The Hamiltonian is given by
\begin{align}
H = \sum_R & \left[ h_{(R,1);(R,2)}(w,u) + h_{(R,2);(R+1,1)}(w,u) \right. \nonumber \\
&\left. + h_{(R,1);(R,1')}(w',u') \right] ~.
\label{eq:Hpertubed}
\end{align}

The Floquet circuit is similarly obtained by applying the two-qubit gates on the three sets of pairs:
\begin{equation}
\label{eq:floq}
{\mathcal U}(\theta, \phi, \theta', \phi') = {\mathcal U}_{\mathrm{even}}(\theta, \phi)\;{\mathcal U}_{\mathrm{odd}}(\theta, \phi) \; {\mathcal U}_{\mathrm{teeth}}(\theta', \phi') ~,
\end{equation}
\begin{align}
    &\mathcal U_{\mathrm{even}}(\theta , \phi )=\prod_R \mathrm{fSim}(\theta , \phi )_{(R,1),(R,2)},\\
    &\mathcal U_{\mathrm{odd}}(\theta , \phi )=\prod_R \mathrm{fSim}(\theta , \phi )_{(R,2),(R+1,1)},\\
&{\mathcal U}_{\mathrm{teeth}}(\theta', \phi') = \prod_R \mathrm{fSim}(\theta', \phi')_{(R,1),(R,1')} ~.
\end{align}
When $w'=0$ ($\theta'=0$ in the Floquet case), the particle number on the original chain is conserved (denoted $N_{1\cup 2}$), and the particle number on each $(R,1')$ site is conserved.
In what follows, we often use more crude grouping of states labelled by sectors $(N_{1\cup 2}, N_{1'} \equiv N - N_{1\cup 2})$.
In this case, in the sector with no particles on the $1'$ sites, the $u'$ term does not operate at all and the Hamiltonian (Floquet circuit) is equivalent to the integrable chain in Eq.~(\ref{eq:Hint}) [Eq.~(\ref{eq:floq0})] for any $u'$ ($\phi'$).

\section{Spectrum  comparative study of the Hamiltonian and Floquet systems}
\label{sec:spectrum}

\begin{figure*}[t]
\centering
\includegraphics[width=\linewidth]{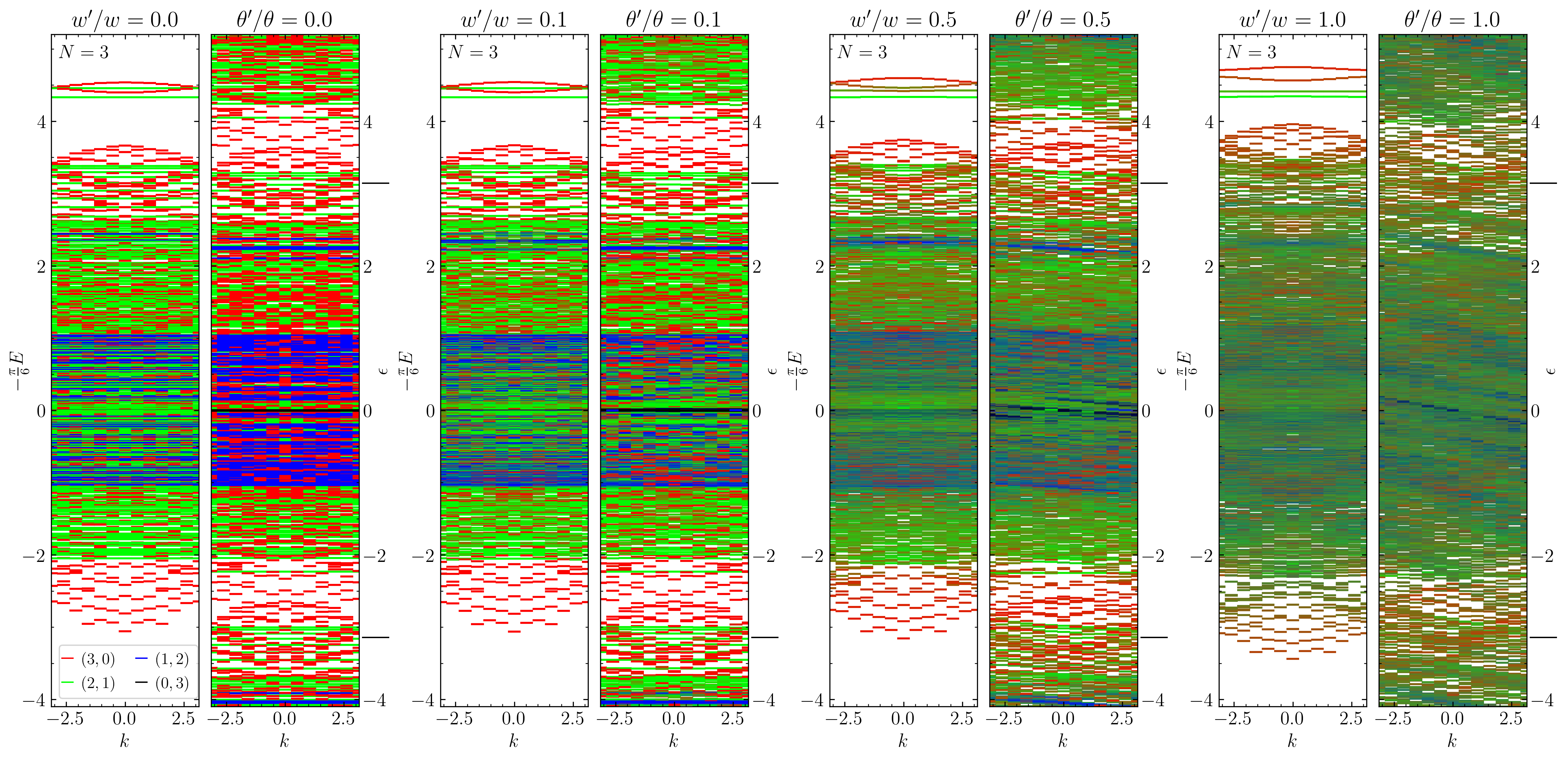}
\caption{Spectrum of the Hamiltonian model (left side in each two-panel group) and Floquet circuit (right side) in the three-particle sector ($N=3$).
The spectra are shown for comparable values of parameters ($w=1$,  $u=u'=4$, $\theta=\pi/6$, $\phi=\phi'=2\pi/3$), and for increasing values of  $w'/w = \theta'/\theta$.
The Hamiltonian energies are multiplied by the time evolution factor of $(-\pi/6)$, and this gives a window wider than $2\pi$ and would require folding to compare with the Floquet quasienergies.
For clarity, instead of such folding of the Hamiltonian spectrum, we show the Floquet circuit quasienergies repeated with period $2\pi$.
The Floquet Brillouin zone $\epsilon \in [-\pi, \pi)$ is marked with grey dashes on the quasienergy scale. 
For easier comparison with the Hamiltonian case, copies of the quasienergy spectrum are shown beyond the first Floquet Brillouin zone.
\red{For the unperturbed models with $w'/w=0$ and $\theta'/\theta=0$, different colors represent states belonging to different $(N_{1\cup 2}, N_{1'})$ sectors, as shown in the legend in the leftmost group.  For the perturbed models,}
the color mixing red/green/blue represents the square of the norms of the projections of each eigenstate in the sectors with $(N_{1\cup 2}, N_{1'}) = (3,0), (2,1), (1,2)$ respectively\red{\sout{ (conserved for $w'/w=\theta'/\theta=0$)}}; specifically, the color in RGB is $(255\cdot P(3,0), 255\cdot P(2,1), 255 \cdot P(1,2))$.
[Note that when $w'/w = \theta'/\theta = 0$ all states in the sector $(0,3)$ have zero energy and their color is RGB (0,0,0), which is black.]
The system size we choose is rather small ($\Luc=12$ unit cells) to avoid overwhelming the plots. }
\label{fig:spectrum3}
\end{figure*}

In this section we present the spectra of the Hamiltonian (\ref{eq:Hpertubed}) and of the Floquet circuit (\ref{eq:floq}) for comparable values of parameters. 
In particular, for the Floquet circuit, we use the same parameters used in the GQAI experiment, namely $\theta=\pi/6$ and $\phi=\phi'=2\pi/3$.
We expect this model to resemble the continuous Hamiltonian evolution for a ratio of parameters $u/w = \phi/\theta = 4$ and $u'/w = \phi'/\theta = 4$.
\red{We will first discuss the general features of the spectrum in the  system with $N=1,2,3$ particles.
While the spectrum for $N=3$ is rather complicated, we will nevertheless show some qualitative, distinctive features through a comparison between the Floquet and the Hamiltonian cases. 
We will then focus on the $N=3$ bound states, by examining a narrower window of the spectrum using some observables that can specifically signal the presence of bound states.}

The spectra as a function of momentum $k$ are plotted in Fig.~\ref{fig:spectra12} in a sector with fixed number of particles $N=1$ and $N=2$: 
for the Hamiltonian $H$ in Eq.~(\ref{eq:Hpertubed}) we plot the spectrum $E(k)$ (rescaled with a factor $-\pi/6$ for comparison); for the Floquet circuit we plot the quasienergies $\epsilon(k)$, defined as the complex phases of the eigenvalues of the Floquet operator in Eq.~(\ref{eq:floq}). 
The spectra are computed for increasing values of the ratio $w'/w = \theta'/\theta$. 

For $w'/w = \theta'/\theta = 0$, the particles  on the $1'$ sites cannot hop, so the particle number on the original chain is conserved, and the particle number on each $(R,1')$ site is conserved. 
In this case, in the sector with no particles on the $1'$ sites, the $u'$ ($\phi'$) term vanishes (acts trivially) and the Hamiltonian (Floquet circuit) is equivalent to the original integrable chain for any $u'$ ($\phi'$).
The eigenstates in this sector are represented in red, and are analyzed in more detail in App.~\ref{app:chiral} for the Floquet case.

For $w'/w = \theta'/\theta \neq 0$ the states of the integrable chain hybridize with the states having non-zero occupation of the $1'$ sites.
As a result, gaps open at $k = \pm \pi$ in the single particle spectrum ($N=1$). 
A rearrangement of the spectrum is observed in the two-particle sector ($N=2$): as the bands in the single-particle spectrum become flatter, some gaps open in the two-particle continuum and some isolated states appear in the gaps.  
The bound states (three isolated bands at the top, corresponding to the three dimer configurations in Fig.~\ref{fig:lattice_and_boundstates}(c)) are still observable in both the Hamiltonian and Floquet case as $w'/w = \theta'/\theta$ is increased from 0 to 1: while they overlap with the two-particle continuum for some values of momentum, for other values they are protected by a gap.

The spectra for $N=1$ and $N=2$ have very similar features in the Hamiltonian and Floquet cases.
The most notable difference is the breaking of time-reversal symmetry in the Floquet case which makes the quasienergy spectrum asymmetric for $k \rightarrow -k$.
When we instead compare the Hamiltonian and the Floquet spectra in the three-particle sector ($N=3$, Fig.~\ref{fig:spectrum3}) we observe a substantial difference:
since the quasienergies are defined modulo $2\pi$, the Floquet spectrum corresponds to a ``folded'' Hamiltonian spectrum. 
As a consequence, the 3-particle bound states, which are gapped and thus stable in the Hamiltonian case, are folded and overlap with the continuum of the Floquet spectrum and therefore they are not protected by a gap. 
The mixing of the bound states with the continuum in the Floquet case can lead to the decay of the bound states. Nevertheless, quantitatively this mixing can be still fairly weak, and the bound states may be visible in the spectrum even for fairly large system sizes.

\begin{figure*}
\centering
\includegraphics[width=\linewidth]{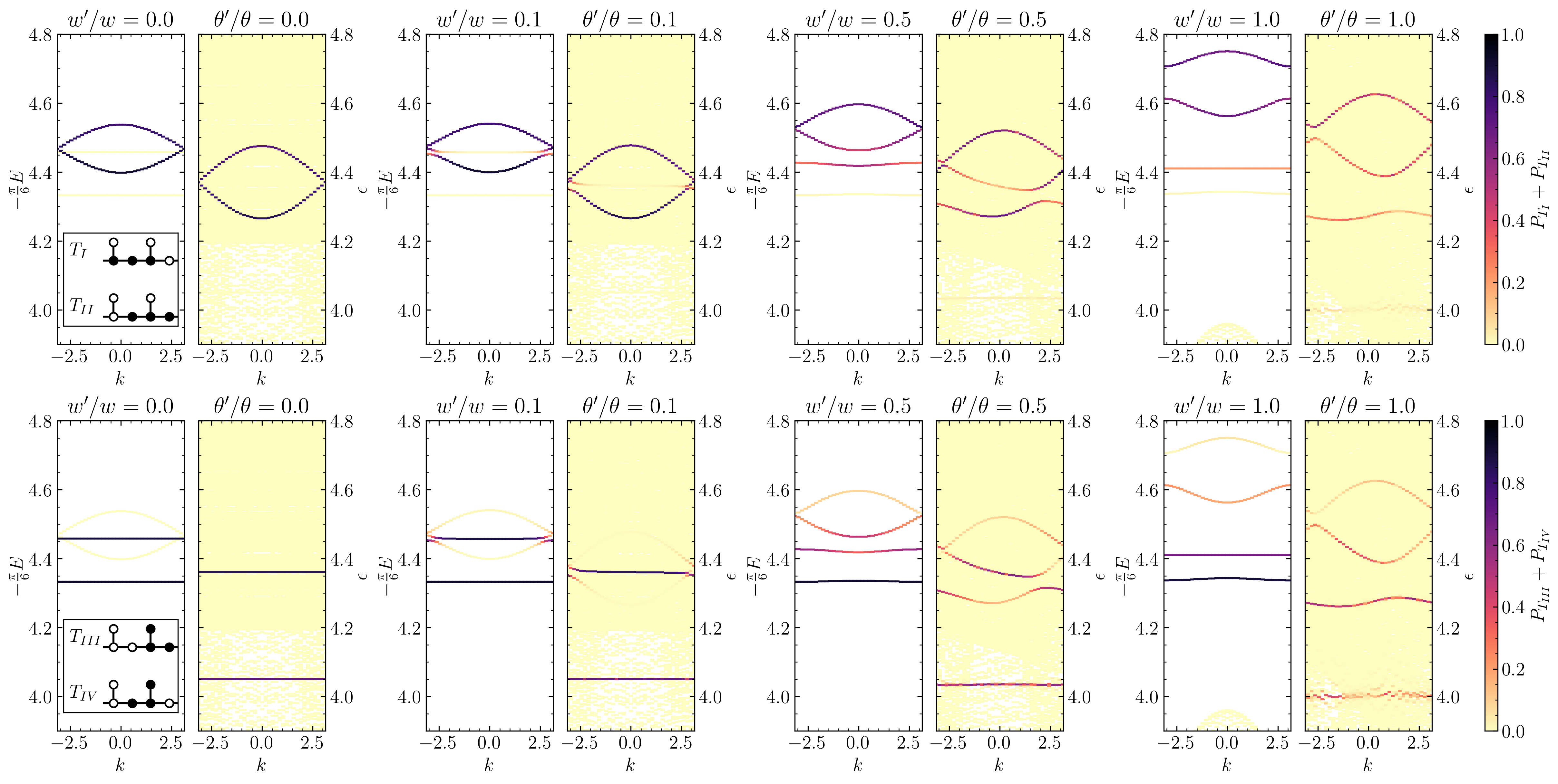}
\caption{Top panels: Probability $P_{T_{\text{I}}} + P_{T_{\text{II}}}$ of having the three particles in consecutive sites along the chain [Eqs.~(\ref{eq:PT1}), (\ref{eq:PT2})]. 
Lower panels: Probability $P_{T_{\text{III}}} + P_{T_{\text{IV}}}$ of having one particle on an extra site and the other two right next to it on the chain [Eqs.~(\ref{eq:PT3}), (\ref{eq:PT4})].
The system size is $\Luc=36$ unit cells.
The system parameters are the same as in Fig.~\ref{fig:spectrum3}.
A narrow window of quasienergies is shown focusing on the three-particle bound states, revealing their character of being primarily chain trimers or chain-extra site trimers and also where significant mixing is present.
In the Hamiltonian system the three-particle bound states are isolated from the continuum and persist in the thermodynamic limit even for $w'/w=1$, while in the Floquet case they are inside a continuum and will decay in the thermodynamic limit but apparently survive to fairly large sizes.
The similarity between the bound states in the Hamiltonian and Floquet systems is notable, allowing to infer properties of the latter ones as well as of the surrounding continuum from the more simple Hamiltonian understanding.
}
\label{fig:overlaps}
\end{figure*}

These bound states are difficult to identify in Fig.~\ref{fig:spectrum3} in the Floquet case, where they overlap with the continuum.
To characterize these states and to discern them from the continuum, it is useful to consider quantities that are sensitive to the relative configuration of the three particles.
Examples of such observables are shown in Fig.~\ref{fig:overlaps}: for each eigenstate $\ket{\psi_{i,k}}$ with momentum $k$, we compute the probabilities of the following configurations for the three particles in neighboring sites [trimers in Fig.~\ref{fig:lattice_and_boundstates}(d)]:
\begin{eqnarray}
P_{T_{\text{I}}} =& \left\lvert \braket{\psi_{i,k}}{\dots \includegraphics[width=0.9cm]{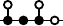} \dots}_{k} \right\rvert^2, \label{eq:PT1} \\
P_{T_{\text{II}}} =& \left\lvert \braket{\psi_{i,k}}{\dots \includegraphics[width=0.9cm]{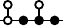} \dots}_{k} \right\rvert^2, \label{eq:PT2} \\
P_{T_{\text{III}}} =& \left\lvert \braket{\psi_{i,k}}{\dots \includegraphics[width=0.9cm]{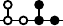} \dots}_{k} \right\rvert^2, \label{eq:PT3}\\
P_{T_{\text{IV}}} =& \left\lvert \braket{\psi_{i,k}}{\dots \includegraphics[width=0.9cm]{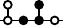} \dots}_{k} \right\rvert^2, \label{eq:PT4}
\end{eqnarray}
where $\ket{\dots}_k$ represents the (normalized) projection of the ``$\dots$'' state in the sector with momentum $k$, i.e.,
\begin{equation}
\label{eq:kstate}
\ket{\dots \includegraphics[width=0.9cm]{bound01.pdf} \dots}_{k}
\equiv \frac{1}{\sqrt{\Luc}} \sum_R e^{ikR} \ket{T_{\text{I}}(R)}.
\end{equation}
An equivalent definition for $P_{T_{\text{I}}}$ (and analogously for the other trimers) is
\begin{equation}
P_{T_{\text{I}}} = \left\lvert \braket{\psi_{i,k}}{\dots \includegraphics[width=0.9cm]{bound01.pdf} \dots}_{k} \right\rvert^2
= \sum_R \left\lvert \braket{\psi_{i,k}}{T_{\text{I}}(R)} \right\rvert^2 ~. 
\end{equation}

In Fig.~\ref{fig:overlaps} we plot the probabilities $P_{T_{\text{I}}} + P_{T_{\text{II}}}$ and $P_{T_{\text{III}}} + P_{T_{\text{IV}}}$ for each eigenstate.
For $w'/w = \theta'/\theta = 0$, we observe two (dispersive) bands with large $P_{T_{\text{I}}} + P_{T_{\text{II}}}$, which correspond to the exact three-particle bound states of the integrable chain (see App.~\ref{app:chiral}).
Two (flat) bands have large $P_{T_{\text{III}}} + P_{T_{\text{IV}}}$: they can be interpreted as bound states of the two particles on the chain localized in the potential of the particle on the extra sites (which acts as an immobile impurity), see App.~\ref{sec:detailed} for detailed study in the Hamiltonian case.
As we turn on the hopping along the teeth ($w'/w = \theta'/\theta \neq 0$), the bands with large $P_{T_{\text{III}}} + P_{T_{\text{IV}}}$ acquire a rather weak dispersion, signaling a very low mobility of the trimers with one particle on the extra sites.

In the cases $w'/w = \theta'/\theta = 0.1, 0.5, 1.0$ the four bound states are still characterized by large values of both $P_{T_{\text{I}}} + P_{T_{\text{II}}}$ and $P_{T_{\text{III}}} + P_{T_{\text{IV}}}$: this shows that even in the Floquet circuit, where the bound states are not protected by a gap, the hybridization is strong among the four bound states but quite weak between the bound states and the continuum.
Even for the largest value we consider ($\theta'/\theta=1.0$), the hybridization with states in the continuum is clearly visible only for one of the four states (the one with quasienergy $\epsilon\sim 4$).

\begin{figure*}
\centering
\includegraphics[width=\linewidth]{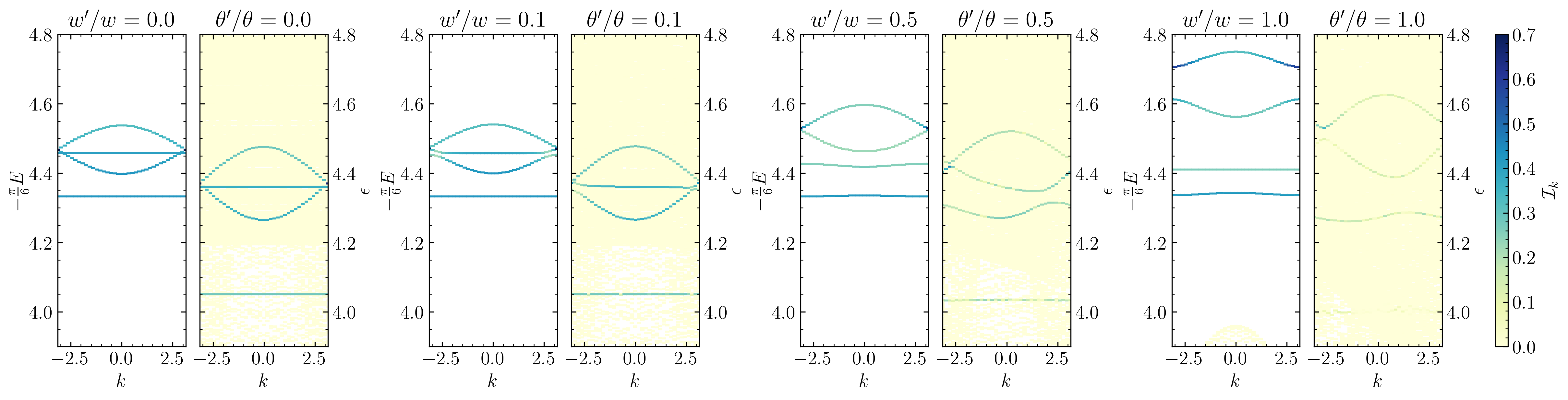}
\caption{Inverse participation ratio (for fixed momentum $k$). 
As explained in the text, this is calculated over basis states used in the momentum-resolved ED, which can be essentially viewed as describing configurations of the particles relative to their center of mass.
The system size is $\Luc=36$ unit cells.
The system parameters and the quasienergy window is the same as in Fig.~\ref{fig:overlaps}.
The IPR detects both types of the three-particle bound states equally well and provides a good measure of the degree of localization of the particles in the bound state.
Detailed study for varying system sizes in the Floquet system at $\theta'=\theta$ (showing the strongest decay of the bound states in this figure) and $k=0$ is performed in Fig.~\ref{fig:IPR}.
}
\label{fig:ipr_spectrum}
\end{figure*}

Another useful quantity to characterize the localization properties of the particles in the bound states is the appropriate inverse participation ratio.
This quantity has been extensively used as a signature of localization induced by disorder both for single particle models \cite{thouless1974electrons,bell1972dynamics} and for many-body systems \cite{Luitz2014,Luitz2015,de2013ergodicity}.
Here, instead, we will use it to probe the localization of particles with respect to their center of mass: 
this localization is induced by interactions and can lead to the presence of bound states.

\red{The idea is to find a quantity that is sensitive also to other ``bound'' configurations of the three particles, that are not captured by the tightest binding configurations defining $P_{T_I}, P_{T_{II}}, P_{T_{III}}, P_{T_{IV}}$.
For example, a bound state can have large overlaps with some other configurations such as $\ket{\dots \includegraphics[width=0.7cm]{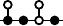} \dots}_{k}$. However, we still expect it to have support on a small (compared to the dimension of the sector) number of such configurations.

To this end, it is useful to define the computational basis in the sector with momentum $k$. Similarly to the definition of Eq.~(\ref{eq:kstate}), each state in the computational basis is defined from a classical configuration $c$ (called {\it representative}) of the three particles as 
\begin{equation}
    \ket{c}_k \equiv \frac{1}{\sqrt{M_c}}\sum_{R=0}^{M_c-1} e^{ikR} \mathcal T^R \ket{c},
\end{equation}
where $\mathcal T$ is the translation operator by one unit cell and $M_c\ge 0$ is the smallest positive integer such that $\mathcal T^{M_c}\ket{c}=\ket{c}$ \footnote{Note that, for the model we are considering, $M_c=\Luc$ for almost all states.}.
The state $\ket{c}_k$ can be also viewed as a normalized projection of the state $\ket{c}$ into the sector with momentum $k$.
The basis is obtained by taking a single representative $c$ for each class of configurations that are related by translations.
}

We define the momentum-resolved \red{inverse} participation ratio (IPR) for a normalized eigenstate $\ket{\psi_{i,k}}$ with momentum $k$ as
\begin{equation}
\mathcal I_k = \sum_c \lvert\braket{\psi_{i,k}}{c}_{k}\rvert^4,
\end{equation}
where the sum runs over all \red{\sout{the possible classes of configurations of three particles related by translation, and $\ket{c}_k$ is a normalized state with momentum $k$ corresponding to such a class.} the representatives (i.e., distinct classes) that define the basis.}
This quantity is an indicator of the localization of the three particles in the frame of their center of mass
\red{\footnote{For states with momentum $k$, it is related to the usual (not momentum-resolved) inverse participation ratio $\mathcal I$ as $\mathcal I_k \approx \Luc \mathcal I$, where the approximation comes from assuming that $M_c=\Luc$ for all configurations.}}.
For a bound state, this quantity converges to a finite value in the large $\Luc$ limit, while it scales as $1/\Luc$ for scattering states of a two-particle bound state with a single particle, and as $1/\Luc^2$ for scattering states of three unbounded particles. 
As shown in Fig.~\ref{fig:ipr_spectrum}, the inverse participation ratio ${\mathcal I}_k$ takes large values for four distinct bands of bound states: these are the same states that were characterized by large values of $P_{T_{\text{I}}} + P_{T_{\text{II}}}$ and $P_{T_{\text{III}}} + P_{T_{\text{IV}}}$ in Fig.~\ref{fig:overlaps}. 
Figure~\ref{fig:ipr_spectrum} shows that, while the value of ${\mathcal I}_k$ remains large for the Hamiltonian case even up to $w'/w=1$, the bound states in the Floquet spectrum exhibit a clear decrease of ${\mathcal I}_k$ as  $\theta'/\theta$ goes from $0.0$ to $1.0$.
This suggests that perturbing the integrable Floquet circuit through the inclusion of additional sites tends to unbind the original bound states of the model.

\section{Eventual instability of the 3-particle bound states in the Floquet model}
\label{sec:instability}

The decrease of the inverse participation ratio as a function of $w'$ shown in the previous section, Fig.~\ref{fig:ipr_spectrum} at fixed $\Luc = 36$, suggests that the localization length in the center of mass frame tends to grow with the perturbation strength.
If this localization length diverges in the thermodynamic limit $\Luc \to \infty$, the 3-particle bound state is unstable. 

\begin{figure}
\centering
\includegraphics[width=\columnwidth]{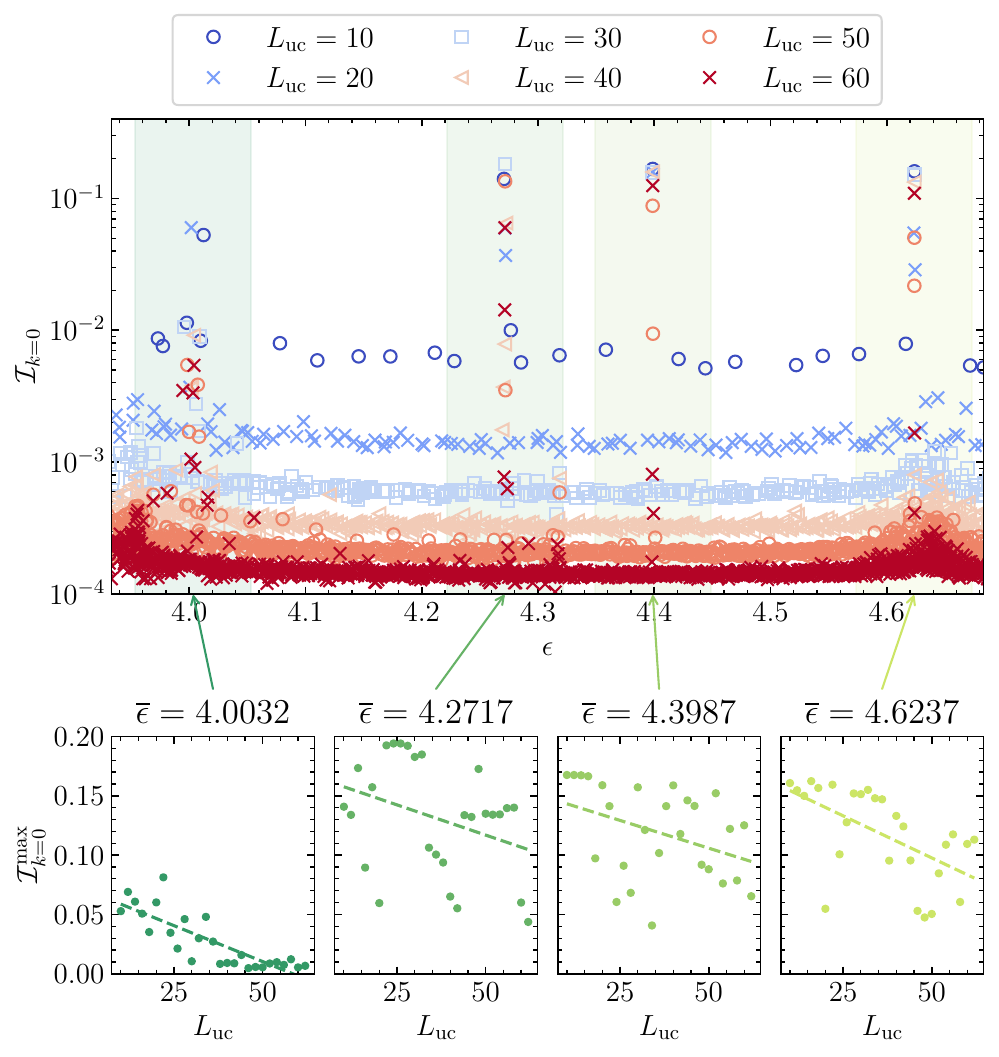}
\caption{Top panel: Inverse participation ratio of Floquet eigenstates with momentum $k=0$ for the system with parameters $\theta=\theta'=\pi/6$, $\phi=\phi'=2\pi/3$. 
The states are organized by the quasienergy on the horizontal axis, and the prominent (approximate) bound states stand out with their large IPR.  Multiple system sizes are shown.
Lower panels: Maximum of $\mathcal I_{k=0}$ in an energy window $[\overline \epsilon-\Delta, \overline \epsilon+\Delta]$ (shaded areas in the top panel), with $\Delta=0.05$, as a function of the system size.
The result of a linear fit, performed to bring out the overall decreasing trend, is shown with a dashed line in each panel.}
\label{fig:IPR}
\end{figure}

In order to probe the eventual decay of the 3-particle bound state, we examine the scaling of the inverse participation ratio with the system size, for $\theta'=\theta$, where the perturbation is of the same order as the unperturbed parameters and the bound states are expected to be least robust of the parameters $\theta' \leq \theta$ considered in the previous sections.
In Fig.~\ref{fig:IPR} (upper panel), we plot the inverse participation ratio of the energy eigenstates in the $k=0$ sector (computed as described in Sec.~\ref{sec:spectrum}) for different numbers of unit cells $\Luc$.
Despite the fairly large perturbation strength, the inverse participation ratio ${\mathcal I}_{k=0}$ still shows four clearly visible peaks that correspond to the four bound states of the unperturbed model.
For large $\Luc$, however, it is possible to notice the effect of the perturbation, which mixes the bound states with the underlying continuum and smears out the peaks.
To quantify this effect, in Fig.~\ref{fig:IPR} (lower panel) we plot the height of the peak (defined as the maximum of ${\mathcal I}_{k=0}$ in an appropriate energy window) as a function of $\Luc$ for the four bound states.
The fluctuations with the system size are still very large, indicating that the results are still very sensitive to the finiteness of the level spacings in the spectrum and to their statistical fluctuations. Nevertheless, all the data shows a decreasing trend with system size, for all the four bound states.
These results suggest that the bound states will ultimately decay in the thermodynamic limit, and that the decay is very slow, leading to persistent 3-particle bound states for numerically and experimentally accessible system sizes.

\begin{figure}
\centering
\includegraphics[width=\linewidth]{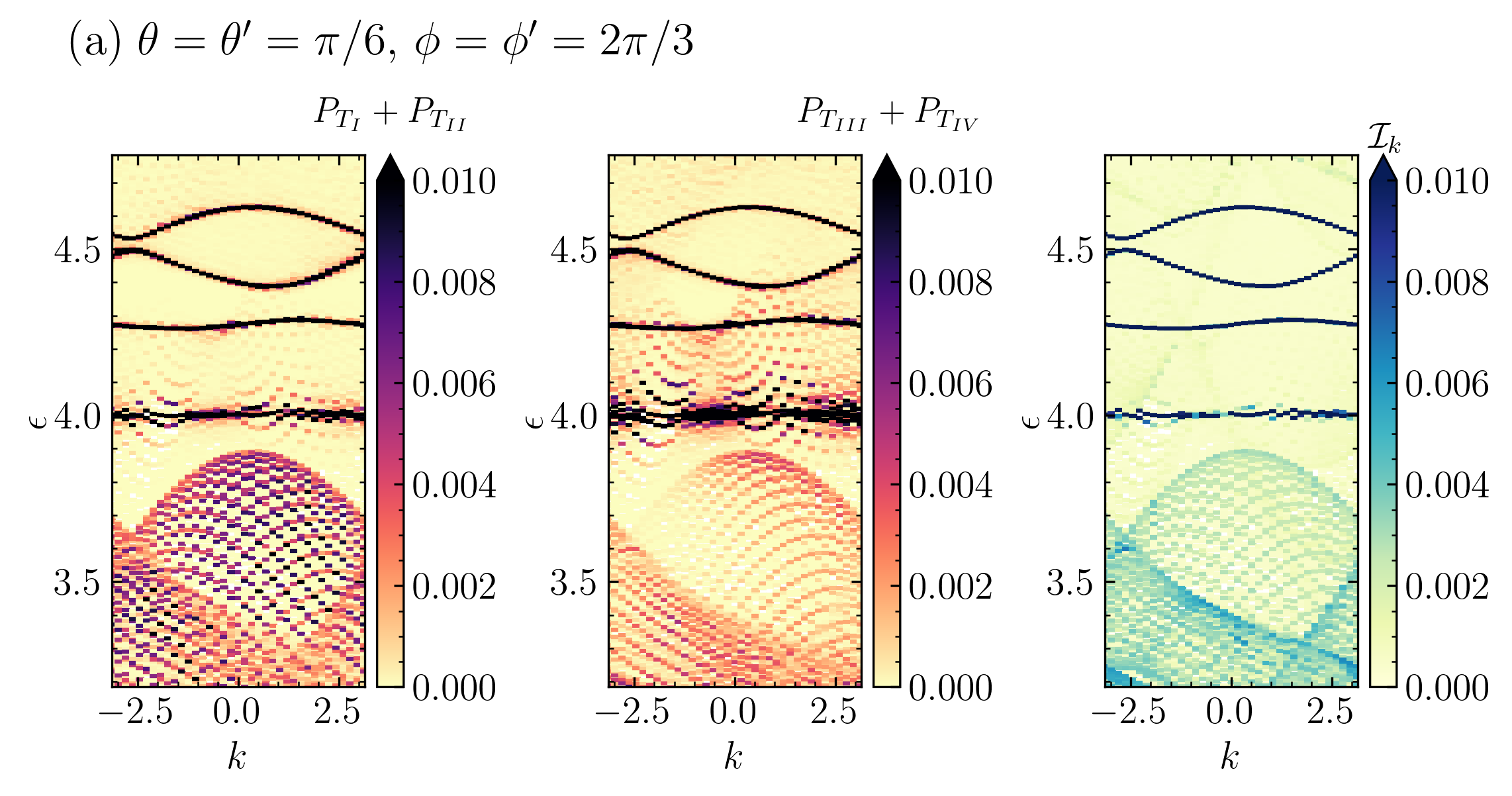}
\includegraphics[width=\linewidth]{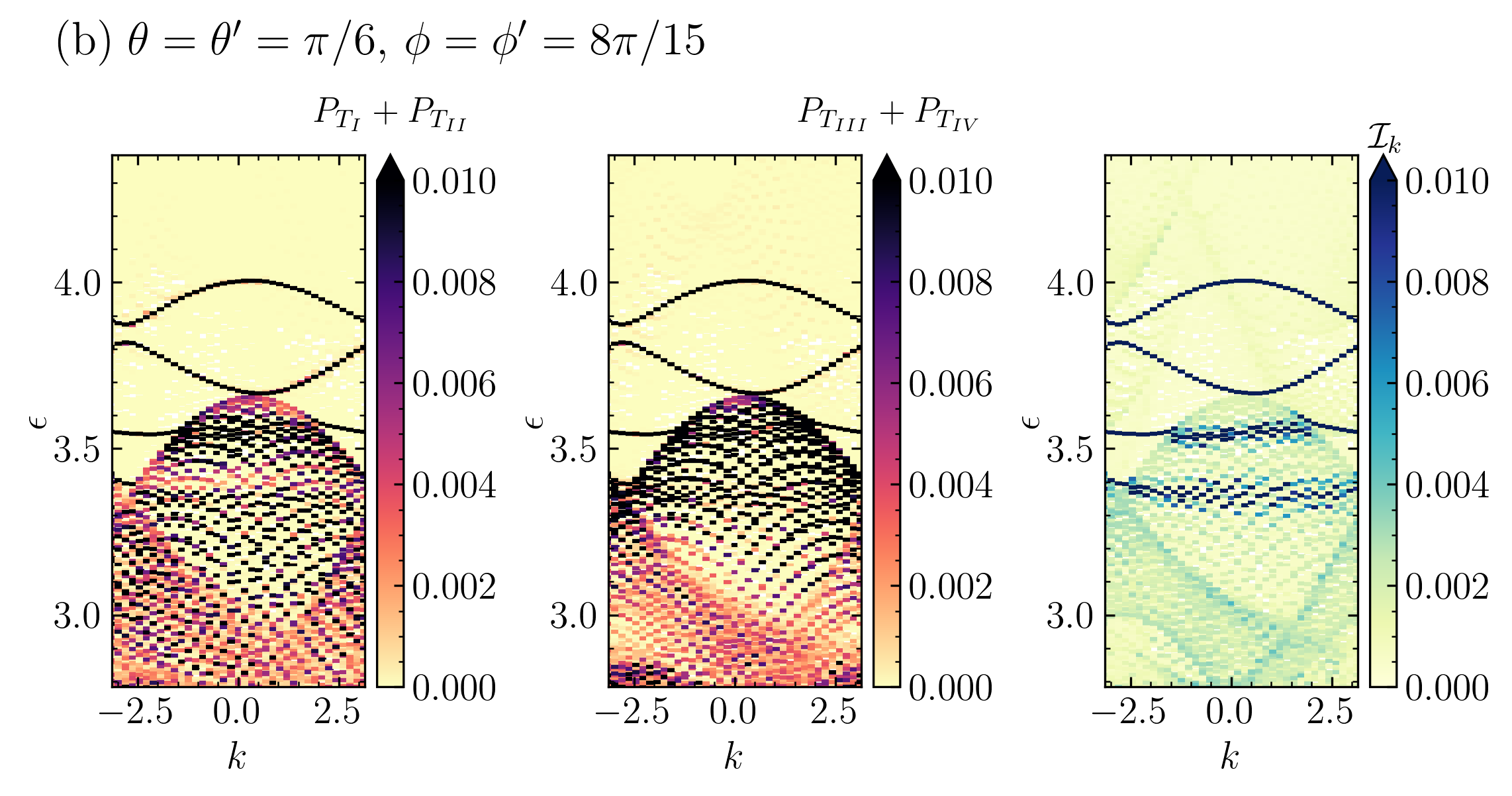}
\caption{Spectrum of the perturbed Floquet circuit focusing on the continuum states near the 3-particle bound states: 
The 2+1 continuum is characterized by larger probability of trimer configurations ($P_{T_{\text{I}}} + P_{T_{\text{II}}}$ and $P_{T_{\text{III}}} + P_{T_{\text{IV}}}$) and larger ${\mathcal I}_k$ than the 1+1+1 continuum. 
Note that the quantities are the same as the ones plotted in Figs.~\ref{fig:overlaps} and \ref{fig:ipr_spectrum} but the color scales are saturated to a maximum value of $0.01$ to bring out the 2+1 states more.
(a) For the choice of parameters considered in the experiment, the bound states are separated in the spectrum from the 2+1 continuum, but one of them is significantly closer. 
(b) For a different choice of parameters, two out of four bound states lie in the 2+1 continuum (for some values of $k$).}
\label{fig:sat}
\end{figure}

\begin{figure}[h]
\centering
\includegraphics[width=\linewidth]{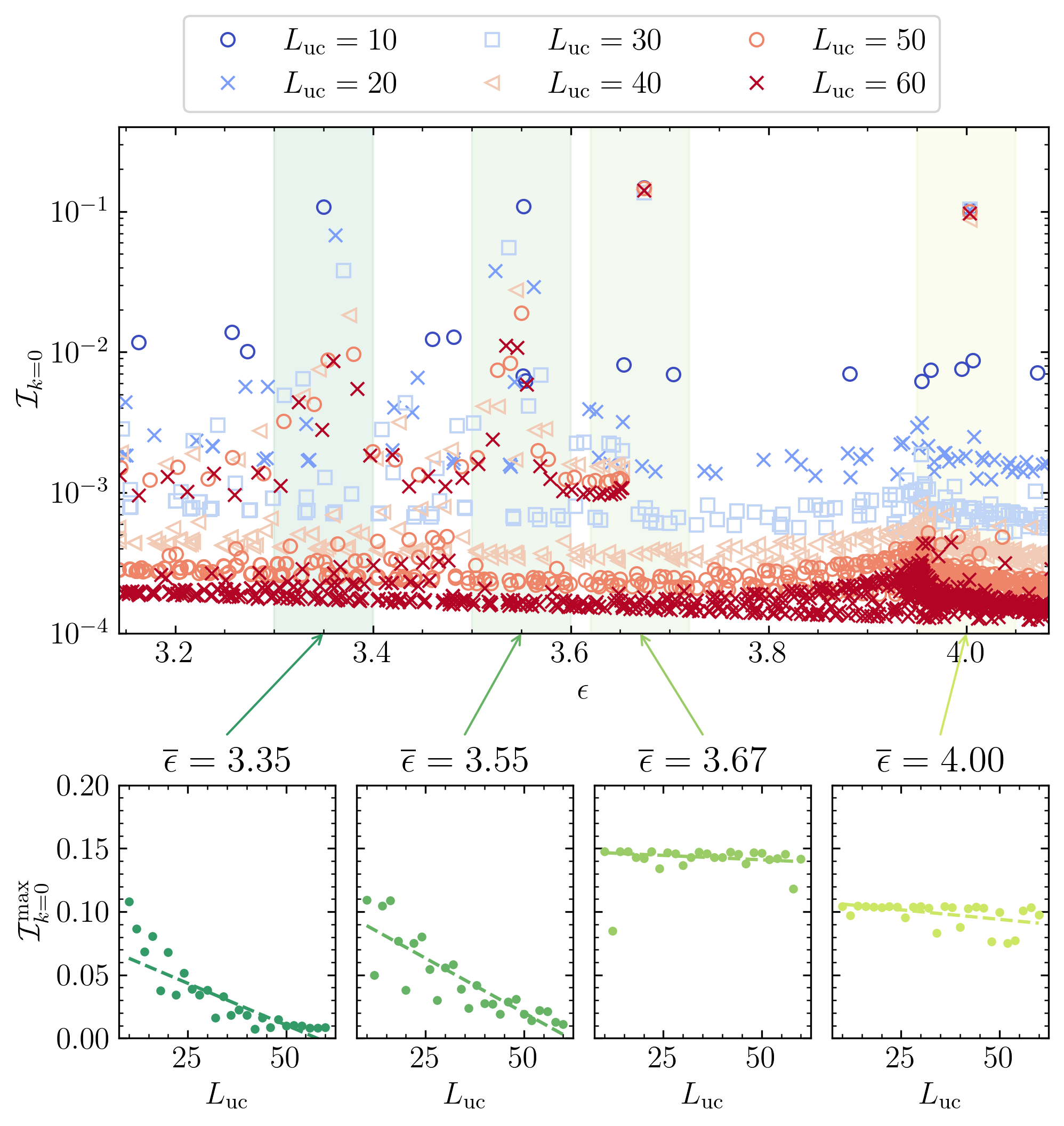}
\caption{\red{Top panel: }Inverse participation ratio of Floquet eigenstates with momentum $k=0$ for the system with parameters $\theta=\theta'=\pi/6$, $\phi=\phi'=8\pi/15$ from Fig.~\ref{fig:sat}(b).
For small system size ($\Luc=10$), four distinct eigenstates stand out with a large value of ${\mathcal I}_{k=0}$.
 \red{Lower panels: Maximum of $\mathcal I_{k=0}$ in an energy window $[\overline \epsilon-\Delta, \overline \epsilon+\Delta]$ (shaded areas in the top panel), with $\Delta=0.05$, as a function of the system size.}
 \red{\sout{Two of them }The two leftmost panels} (the ones \red{which correspond to the eigenstates} overlapping with the 2+1 continuum) show a clear decay of the height of the peak for increasing system size.
The other two show \red{\sout{no apparent }an extremely weak} decay with $\Luc$.
 }
\label{fig:IPRothervalue}
\end{figure}

Of the four bound states, we observe that one of them (the one with quasienergy $\epsilon \sim 4.0032$) exhibits significantly faster decay of ${\mathcal I}_{k=0}$ with the system size (Fig.~\ref{fig:IPR}, lower panels).
We will now argue using numerical experiments that this faster decay is caused by the proximity in the spectrum with scattering states of a two-particle bound state with a single particle (which we will refer to as ``2+1 continuum'' below).
Note that strictly speaking this label refers to states present in the integrable model at $\phi'=0$ in the sector $(N_{1\cup 2}, N_{1'}) = (3, 0)$, while the states highlighted in Fig.~\ref{fig:sat} at $\phi'=\phi$ are their descendants.
As shown in Fig.~\ref{fig:sat}(a), the 2+1 continuum is characterized by a larger probability of trimer configurations and larger ${\mathcal I}_{k}$ compared to the scattering states of individual particles (which we will call ``1+1+1 continuum''). 
The edge of the 2+1 continuum is very close to the lowest bound state, suggesting that these states are responsible for the larger hybridization.
In Fig.~\ref{fig:sat}(b) we also consider a different choice of parameters, for which the spectra of the two lowest bound states are partially enclosed in the 2+1 continuum. In this case, these two states show a fast decay of $\mathcal I_{k=0}$ with the system size $\Luc$, while the other bound states are more resilient (Fig. \ref{fig:IPRothervalue}).

These results suggest that the decay is fast when the 3-particle bound state can decay in the 2+1 continuum, but is very slow when it can only decay in the 1+1+1 continuum. This explains the apparent robustness observed in the experiment \cite{morvan2022formation}.

\section{Perturbative analysis of eventual instability}
\label{sec:perturbative}

To address the question of an eventual instability, it is useful to study the matrix elements of the perturbation (i.e., the hopping along the teeth) in the basis of the unperturbed eigenstates of the Floquet model with $\phi'=\phi$, $\theta'=0$.
The perturbation of the Floquet circuit is
\begin{equation}
{\mathcal U}^{-1}_{\mathrm{teeth}}(0, \phi') \; {\mathcal U}_{\mathrm{teeth}}(\theta', \phi') = \exp(-i \theta' V) \approx \mathbb{1} - i \theta' V ~,
\end{equation}
where $\theta'\ll 1$ is the small perturbative parameter, and $V$ is defined as
\begin{equation}
V = -\sum_R (\ketbra{10}{01}_{(R,1),(R,1')}+\text{H.c.}).
\end{equation}
It is adequate for our purposes to use intuition from the Hermitean perturbation theory treating the Floquet quasienergies as the unperturbed energies and $\theta' \cdot V$ as the perturbation.

We label the four bound states of the $\theta'=0$ model in the sector with total momentum $k=0$ as $\ket{\psi_n}$ with $n=0,1,2,3$:
$\ket{\psi_0}$, $\ket{\psi_2}$ have quasienergies $\epsilon_0 \approx 4.0512$ and $\epsilon_2 \approx 4.3614$, respectively, and belong to the sector with $(N_{1\cup 2}, N_{1'}) = (2,1)$;
$\ket{\psi_1}$, $\ket{\psi_3}$ have quasienergies $\epsilon_1 \approx 4.2657$ and $\epsilon_3 \approx 4.4755$, and belong to the sector with $(N_{1\cup 2}, N_{1'}) = (3,0)$.
We generally expect a bound state $\ket{\psi_n}$ to be unstable to a perturbation $V$ if the ``Fermi's golden rule rate'' $\Gamma_n = 2\pi (\theta')^2 \sum_j |\braket{\psi_n|V}{\epsilon_j}|^2 \delta(\epsilon_j-E_n)$ is finite, where  $V_{nj} \equiv \braket{\psi_n|V}{\epsilon_j}$ is the matrix element connecting the $n$-th bound state with the state $\ket{\epsilon_j}$ in the continuum and $E_n$ is the energy of the bound state.

From the numerical study we know that the 3-particle bound states of interest to us do not overlap in energy with the 2+1 states (i.e., scattering states of a 2-particle bound state and a particle).  Then
conservation of energy and momentum implies that the bound states can only decay in the 1+1+1 continuum for $k=0$.  For these continuum states, from a simple counting argument we expect a density of states $\propto \Luc^2$. 
The matrix element $V_{nj}=\braket{\psi_n|V}{\epsilon_j}$ between a state in the 1+1+1 continuum and a bound state (i.e., a localized state in the center of mass frame) can similarly be estimated from a simple argument: the state $\ket{\epsilon_j}$ of the three particles can be approximated as a (properly symmetrized) product of three plane waves, while $\ket{\psi_n}$ is a single plane wave (with $k=0$); the matrix element is non-zero only when the three particles are next to each other; taking into account the normalizations of the plane waves, we get that the  matrix element scales as $|V_{nj}|\propto \Luc \cdot (1/\sqrt{\Luc}) (1/\sqrt{\Luc})^3=\Luc^{-1}$. 
We then expect the product between the density of states and $|V_{nj}|^2$ to yield an $O(1)$ rate in the thermodynamic limit. 

Note that a similar argument for a decay into the 2+1 continuum would give a density of states $\propto \Luc$ and a matrix element $|V_{nj}|\propto \Luc \cdot (1/\sqrt{\Luc}) (1/\sqrt{\Luc})^2 = \Luc^{-1/2}$, resulting, again, in a finite rate in the thermodynamic limit.

\begin{figure}[h]
\centering
\includegraphics[width=\linewidth]{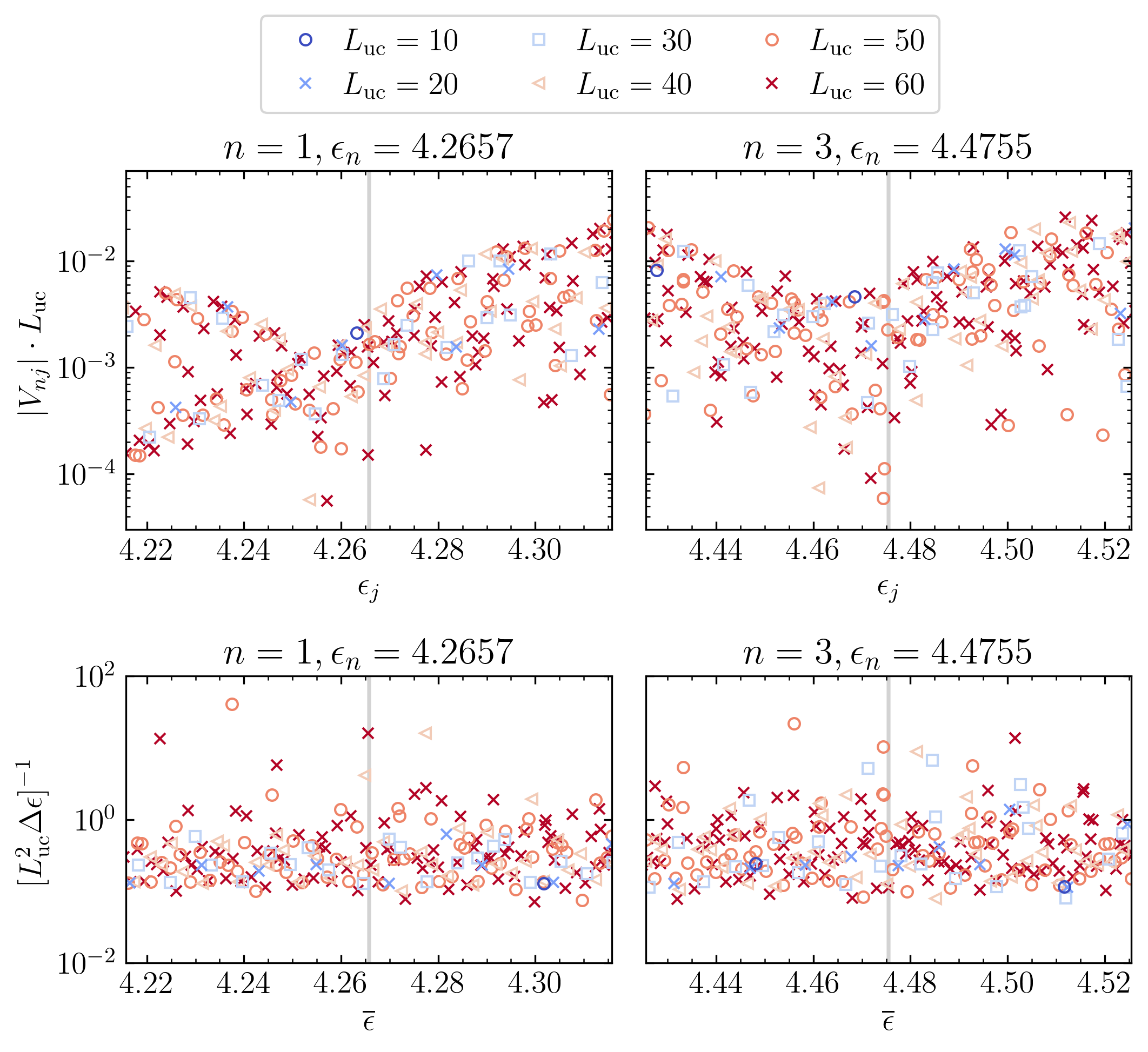}
\caption{Top: absolute value of the matrix elements $V_{nj}$ (connecting the bound state $\ket{\psi_n}$ with the eigenstate $\epsilon_j$) multiplied by the system size $\Luc$, as a function of the quasienergy $\epsilon_j$. 
The data collapse shows that $|V_{nj}|\propto \Luc^{-1}$ consistent with 1+1+1 continuum.
Bottom: inverse level spacing of neighboring pairs of Floquet eigenstates $(\Delta \epsilon)^{-1}$ multiplied by $\Luc^{-2}$, as a function of the average quasienergy of the pair. Only states that have a non-zero matrix elements with $\ket{\psi_n}$ are considered. 
The data collapse shows the predicted scaling for the density of states $(\Delta \epsilon)^{-1} \propto \Luc^2$. 
The vertical grey lines indicate the energy $\epsilon_n$ of the bound state.
The Floquet system parameters are $\phi=\phi'=2\pi/3$, $\theta=\pi/6$, $\theta'=0$, i.e., the ``integrable point''.
The data are plotted for the bound states with $n=1$ (left) and $n=3$ (right), which belong to the sector with $(N_{1\cup 2}, N_{1'}) = (3,0)$.}
\label{fig:dos2}
\end{figure}

\begin{figure}[h]
\centering
\includegraphics[width=\linewidth]{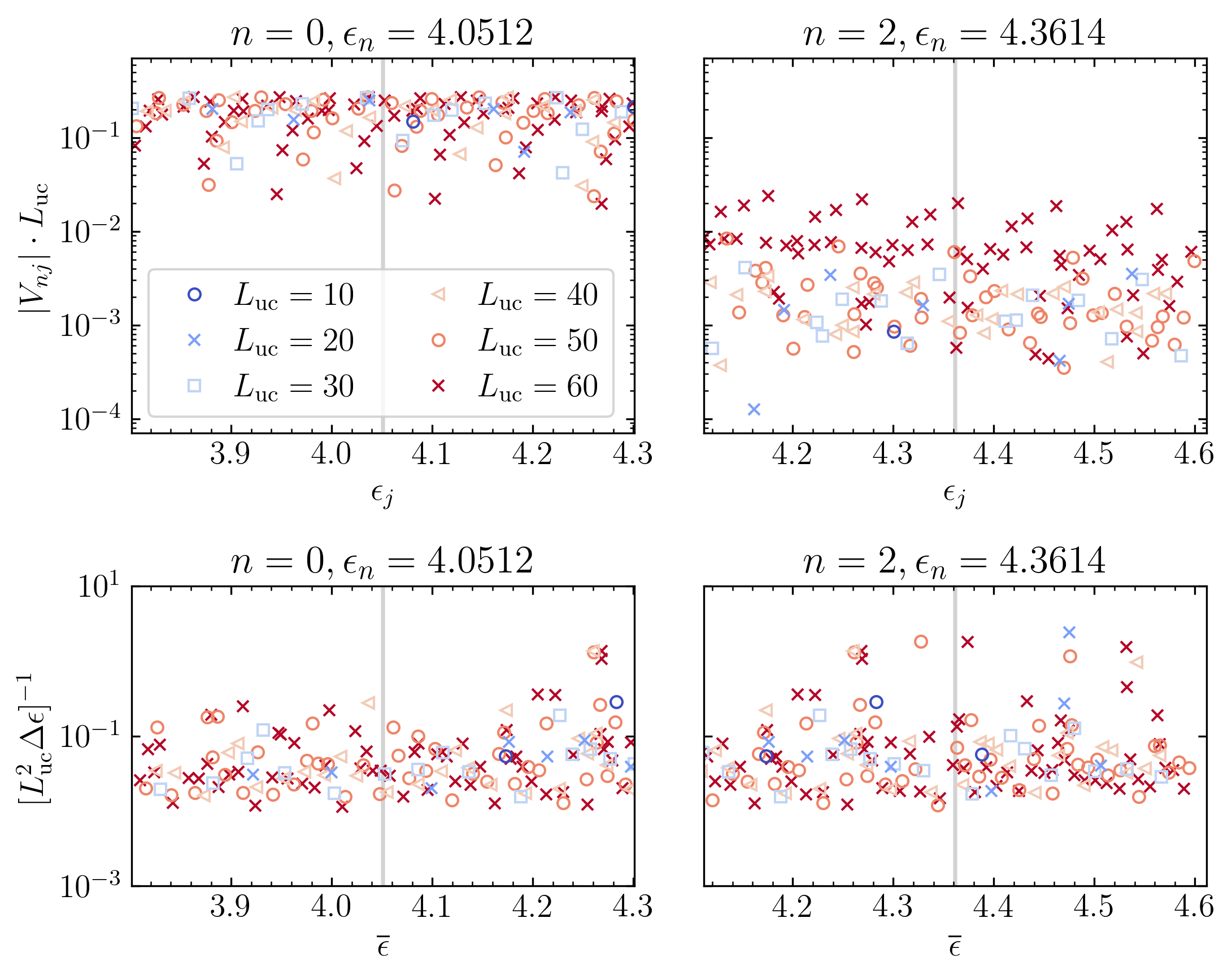}
\caption{ Similar analysis as in Fig.~\ref{fig:dos2}, but for the bound states $n=0$ (left) and $n=2$ (right), which belong to the sector with $(N_{1\cup 2}, N_{1'}) = (2,1)$.
In the case of degeneracies, we consider a single $\ket{\epsilon_j}$, proportional to the projection of $V\ket{\psi_n}$ on the degenerate subspace.
The data collapse shows that both the matrix elements (top) and the density of states (bottom) obey the predicted scaling with the system size for 1+1+1 continuum.}
\label{fig:dos1}
\end{figure}

In Figs.~\ref{fig:dos2} and \ref{fig:dos1}, we numerically check our prediction for the scaling of the matrix elements and of the density of states for the decay into the 1+1+1 continuum. 
We plot the matrix elements $|V_{nj}|$ (including only the ones that are non-zero)  multiplied by $\Luc$. 
As a measure of the density of states, we consider pairs $\epsilon_j, \epsilon_{j+1}$ of nearby levels in the quasienergy spectrum (including only the states with non zero $|V_{nj}|$) and we plot the inverse spacing $(\Delta \epsilon)^{-1}=(\epsilon_{j+1}-\epsilon_j)^{-1}$  multiplied by $\Luc^{-2}$ as a function of the average quasienergy $\overline{\epsilon}=(\epsilon_{j+1}+\epsilon_j)/2$. We find that, for all bound states $n=0,1,2,3$, both $|V_{nj}|\cdot \Luc$ and $(\Luc^2 \Delta \epsilon)^{-1}$ show a good data collapse, with no systematic dependence on system size \footnote{For $n=2$, we see that $|V_{nj}|\cdot \Luc$ takes systematically larger values for $\Luc=60$: as we discuss in App.~\ref{app:overview}, the reason is the hybridization of the $n=2$ bound state in the $(N_{1\cup 2}, N_{1'}) = (2,1)$ sector (which is not protected by a gap nor by integrability) with states in the continuum belonging to the same sector.  The hybridization is especially large for $\Luc=60$, being an outlier in our data in Fig.~\ref{fig:IPR_resolved}.}, confirming our predictions that $|V_{nj}|\propto \Luc^{-1}$ and $(\Delta \epsilon)^{-1}\propto \Luc^2$.

\begin{figure}[h]
\centering
\includegraphics[width=\linewidth]{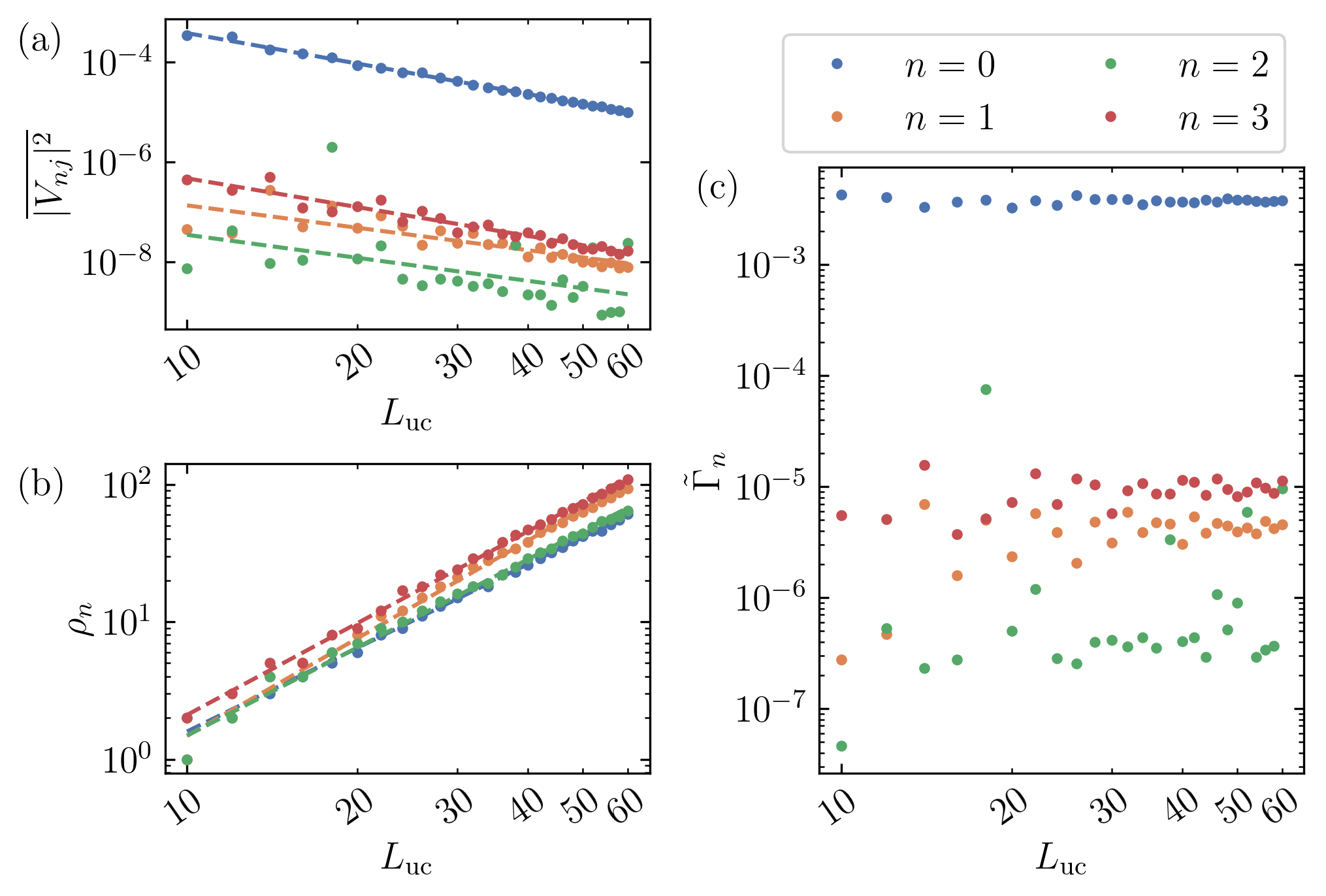}
\caption{(a) Average of $|V_{nj}|^2$ as a function of the system size $\Luc$ for $n=0,1,2,3$ and $\epsilon_j$ belonging to the energy windows $\epsilon_n \pm \Delta_n$ of Figs.~\ref{fig:dos1} and \ref{fig:dos2}. 
The dashed lines show the results of linear fits (in log-log scale) $\overline{|V_{nj}|^2}\sim k_n \Luc^{\alpha_n}$, with $\alpha_n = (-2.03, -1.49, -1.53, -1.91)$ and $k_n = (0.04, 4\cdot 10^{-6}, 1\cdot 10^{-6}, 4\cdot 10^{-5})$ for $n=0,1,2,3$ respectively.
(b) Density of states $\rho_n$ estimated as the number of eigenstates in the energy window $\epsilon_n \pm \Delta_n$ divided by the total width $2\Delta_n$, as a function of system size. 
Only states with non-zero matrix elements are included. 
The results of the fits $\rho_n \sim c_n \Luc^{\beta_n}$ have parameters $\beta_n = (2.02, 2.36, 2.13, 2.22)$ and $c_n = (0.015, 0.006, 0.011, 0.013)$. 
(c) Decay rate $\tilde{\Gamma}_n = 2\pi \rho_n \overline{|V_{nj}|^2}$.}
\label{fig:scaling_rate}
\end{figure}

The scaling is further analysed in Fig.~\ref{fig:scaling_rate}(a) by plotting the average of $|V_{nj}|^2$ for each $n=0,1,2,3$ over the energy windows plotted in Figs.~\ref{fig:dos1} and \ref{fig:dos2}, as a function of system size.
The results of the fits show a dependence $\overline{|V_{nj}|^2}\propto \Luc^\alpha$ with $\alpha$ in the range $-2.03, -1.49$, in rough agreement with the expect scaling $|V_{nj}|^2 \sim \Luc^{-2}$ for matrix elements to the 1+1+1 continuum states. 
In Fig.~\ref{fig:scaling_rate}(b) we plot the density of states $\rho_n$ (computed as the number of states in the same energy windows, divided by the width of the windows) as a function of $\Luc$: the results agree with the expected scaling $\rho_n\propto \Luc^2$.
The rate of decay is then computed as \red{\sout{$\Gamma_n = (\theta')^2 2\pi \rho_n \overline{|V_{nj}|^2} \equiv (\theta')^2 \tilde{\Gamma}_n$}}
\red{\begin{equation}
    \Gamma_n = (\theta')^2 2\pi \rho_n \overline{|V_{nj}|^2} \equiv (\theta')^2 \tilde{\Gamma}_n
\end{equation}}
and $\tilde{\Gamma}_n$ is plotted in Fig.~\ref{fig:scaling_rate}(c). 
The rate is approximately constant, with no evident dependence on system size (except for a mild increase, which is visibly present only for small $\Luc$). 
However, the values of the rates ${\tilde \Gamma}_n$ for different bound states span many orders of magnitudes.
For $n=1,2,3$ they are smaller than $10^{-5}$, suggesting that the bound states may persist at finite size for large values of the perturbation (or equivalently, for large system size with a fixed perturbation strength).
[The corresponding physical decay rates $\Gamma_{1,2,3} < 10^{-5} \times (\theta')^2$, for smallish $\theta' \leq \pi/6$, give lifetimes that can exceed $10^5$ Floquet cycles.]
The $n=0$ has the largest decay rate ($\Gamma_0 \approx 4\cdot 10^{-3} \times (\theta')^2$, significantly larger than the other three), and hence it should exhibit a more enhanced decay with system size for $\theta'\neq 0$, in agreement with the results of Fig.~\ref{fig:IPR}.
We remark, however, that the four bound states studied in Sec.~\ref{sec:instability} for $\theta'=\theta$ cannot be simply attributed in a one-to-one correspondence to the unperturbed states $\ket{\psi_n}$, with $n=0,1,2,3$ at $\theta'=0$, because of the strong hybridization of the bands visible at the studied $\theta' \neq 0$, in particular for $n=1,2,3$; hence one should not use such perturbative estimates literally for all $\theta'$ of interest.

\red{Nevertheless, it is suggestive, to use the estimate of the decay rates to interpret the decrease of the IPR peaks of Fig.~\ref{fig:IPR} with increasing $\Luc$.
The Wigner-Weisskopf theory for the decay of a state in quantum-mechanics relates the decay rate with the width of the resonance in the frequency domain.
At finite system size, this treatment can break down because the spectrum is discrete: in order to observe the finite width of a resonance, we need the level spacing to be much smaller than the width. 
Extrapolating our estimates of $\Gamma_n$ to the value $\theta'=\theta=\pi/6$ of Fig.~\ref{fig:IPR} (well beyond the perturbative regime), we find $\Gamma_0 \approx 10^{-3}$ and $\Gamma_{1,2,3} < 3 \cdot 10^{-6}$.
From our fits for the density of states close to the four peaks for $\theta'=\theta=\pi/6$ we  are finding that $\rho_n \approx f_n \Luc^2$ with $f_n\approx (0.04, 0.04, 0.03, 0.02)$, so the average level spacing $\rho_n^{-1}$ becomes of the order of $\Gamma_n$ for $\Luc \approx (\Gamma_n f_n)^{-1/2}$, which results in $\Luc \approx 150$ for $n=0$ and $\Luc >2700$ for $n=1,2,3$.
This estimate confirms that larger system sizes are needed in order to clearly observe the decay for the three rightmost peaks in the inverse participation ratio in Fig.~\ref{fig:IPR}.
If the system size is not sufficiently large, this measure is sensitive to the fluctuations of the individual eigenergies, resulting in the noisy dependence observed in Fig.~\ref{fig:IPR}.}

\section{Conclusions}
\label{sec:conclusions}

In this work, we analysed the robustness of the bound states observed in the Google Quantum AI experiment~\cite{morvan2022formation}.
We compared the Hamiltonian and the Floquet spectrum, showing that the bound states in the $N=3$ sector, which are protected by a gap in the Hamiltonian case, overlap (fold) with the other edge of the spectrum when trying to connect to the Floquet case.
This is consistent with the direct study in the Floquet case where the bound states are surrounded by continuum states. 
We characterized the bound states by studying their overlaps with the trimer configurations and their inverse participation ratio resolved in sectors of total momentum.
Our results suggest that similar many-body spectroscopic techniques as the ones applied to observe the bands of exact bound states in the integrable circuit can be used to detect the bands of approximate bound states in the perturbed circuit and to measure properties such as the maximal band velocity and their microscopic structure.
For example, depending on the band, the particles in the bound state either reside primarily on the chain as in the trimers $T_{\text{I}}$ and $T_{\text{II}}$ in Fig.~\ref{fig:lattice_and_boundstates}, or have one particle on the extra sites as in the trimers $T_{\text{III}}$ and $T_{\text{IV}}$ in the same figure.
Such more detailed dynamical and structural properties of the bound states in fact change significantly as one varies $\theta'$ from $0$ to $\theta$ (without much effect on their apparent robustness, in part because the bound states primarily mix among themselves) and could be probed directly in experiments.

For $\theta'=0$, particles located on the extra sites act as impurities in an integrable model on the chain. 
Recently, exact spatially bound states  inside a continuous spectrum have been proposed in some integrable Hamiltonian models in the presence of an impurity \cite{Zhang_BoundStatesHubbard,sugimoto2023many}.
Interestingly, one of the 3-particle bound states in our modelling of the experimental system at $\theta'=0$, namely $n=2$ state with $\epsilon \approx 4.3614$ in Sec.~\ref{sec:perturbative} from the sector $(N_{1\cup 2}, N_{1^\prime}) = (2, 1)$, appears to correspond to a similar instance in the Floquet setting, see discussion in App.~\ref{app:overview} (where this state is referred to as $\epsilon \approx -1.92$ from $2\pi$ shift).
It would be interesting to investigate  possible existence and stability of such states more broadly in the Floquet XXZ model with an impurity, both theoretically and experimentally.

For $\theta' \neq 0$, while the bands of bound states are clearly visible even for fairly large system sizes, our finite size scaling analysis shows that the bound states tend to decay for increasing $\Luc$.
The decay is more rapid for one of the four bound states, due to the proximity with the 2+1 continuum.
For other values of the parameters, other bound states can become similarly more unstable: we anticipate that such a difference in the robustness of the bound states can be probed in the same experimental apparatus, by preparing different initial states [such as the trimer configurations $T_{\text{III}}$ and $T_{\text{IV}}$ in Fig.~\ref{fig:lattice_and_boundstates} that would decay much faster for modified parameters as in Figs.~\ref{fig:sat}(b) and \ref{fig:IPRothervalue}].
A numerical analysis of the matrix elements and of the density of states in the unperturbed model confirms the presence of small but finite decay rates for all the bound states.

Our explanation for the current experiments on the non-integrable model is thus a quantitative few-body one and does not require true many-body unusual thermalization.
An interesting question for future work is the possibility of so-called weak integrability-breaking perturbations for the Floquet XXZ model, and for Floquet integrable models in general.
These perturbations, which can be systematically constructed to preserve integrability up to a given order in the perturbation strength, have been studied in the context of Hamiltonian integrable models.
Understanding their possible structure in Floquet circuits would allow for the experimental verification of slow dynamics in digital quantum devices, such as the one used in the Google Quantum AI experiment. 

{\it Note added - } A study by Hudomal {\it et.~al.}~\cite{Hudomal}, which appeared shortly after our work, also studies the bound states and the spectral properties of the GQAI experiment.  Reference~\cite{Hudomal} focuses on different properties of the spectrum and studies the time evolution using TEBD simulations, and reaches different conclusions about the robustness of the bound states in the infinite time and infinite system size limit, which however may be hindered by their available time scales/sizes compared to potentially very long decay times.

\begin{acknowledgements}
We thank Jason Alicea, Sanjay Moudgalya, Balázs Pozsgay, and Pablo Sala for insightful discussions.
FMS acknowledges support provided by the U.S.\ Department of Energy Office of Science, Office of Advanced Scientific Computing Research, (DE-SC0020290), by Amazon Web Services, AWS Quantum Program, and by the DOE QuantISED program through the theory consortium ``Intersections of QIS and Theoretical Particle Physics'' at Fermilab.
OIM acknowledges support by the National Science Foundation through grant DMR-2001186.
A part of this work was done at the Aspen Center for Physics, which is supported by National Science Foundation grant PHY-2210452. 
This work was partially supported by a grant from the Simons Foundation.
\end{acknowledgements}

\bibliography{bib}

\appendix

\section{Floquet XXZ and gap in the chiral spectrum}
\label{app:chiral}
The unperturbed Floquet circuit in Eq.~(\ref{eq:floq0}) has a simple brickwall structure, and can be conveniently diagonalized by considering the operator 
\begin{equation}
{\mathcal V}_0(\theta, \phi)  \equiv {\mathcal U}_{\text{even}}(\theta, \phi)\; {\mathcal T} ,
\end{equation}
where ${\mathcal T}$ is the translation operator by one chain site \cite{Aleiner2021}. 
We can then write the Floquet operator as
\begin{equation}
{\mathcal U}_0 = {\mathcal U}_{\mathrm{even}}\; {\mathcal U}_{\mathrm{odd}}={\mathcal U}_{\mathrm{even}}\; \mathcal T\;{\mathcal U}_{\mathrm{even}}\; \mathcal T^{-1}=\mathcal V_0^{2}\; \mathcal T^{-2} ,
\end{equation}
where we have dropped the $(\theta, \phi)$ dependence for easier readability.
We can diagonalize ${\mathcal V}_0$ in a fixed momentum sector with respect to the translation of two chain sites (which corresponds to one lattice unit used throughout the text), i.e., we find simultaneous eigenvalues and eigenvectors of ${\mathcal V}_0$ and ${\mathcal T}^2$:
\begin{equation}
{\mathcal T}^2 \ket{\gamma,k} = e^{ik} \ket{\gamma,k}, \qquad {\mathcal V}_0 \ket{\gamma,k} = e^{i\gamma} \ket{\gamma,k}.
\end{equation}
(Note that the momentum $k$ has exactly the same meaning as in the main text, since ${\mathcal T}^2$ corresponds to translation by one unit cell.)
Then, the states $\ket{\gamma,k}$ are eigenstates of the Floquet operator ${\mathcal U}_0$ with quasienergies
\begin{equation}
\epsilon = 2\gamma - k ~~(\text{mod}~~ 2\pi) ~.
\end{equation}

The model was solved in Ref.~\cite{Aleiner2021} using Bethe ansatz, and the following dispersion relation of the $\ell$-particle bound state was found for generic $\ell$
\begin{equation}
\cos\left(\epsilon_{\ell\text{-string}}(k)-\chi\right)=\cos^2(\alpha)-\sin^2(\alpha) \cos(k),
\label{eq:eellstring_Floquet}
\end{equation}
where
\begin{align}
& \chi = \ell \phi -2\arctan\left(\frac{\tan(\phi/2)\tanh(\eta)}{\tanh(\ell \eta)}\right), \\
& \cos^2(\alpha) = \frac{\cos^2(\theta) \sinh^2(\ell \eta) }{\cos^2(\theta) \sinh^2(\ell \eta) + \sin^2(\theta) \sinh^2(\eta)}, \\
& \sinh^2\left(\eta\right) = \frac{\cos^2(\theta)-\cos^2(\phi/2)}{\sin^2(\theta)}.
\end{align}
Our numerical study below of the spectra of ${\mathcal U}_0$ and the bound states is consistent with these predictions.

\begin{figure}[h]
\centering
\includegraphics[width=\linewidth]{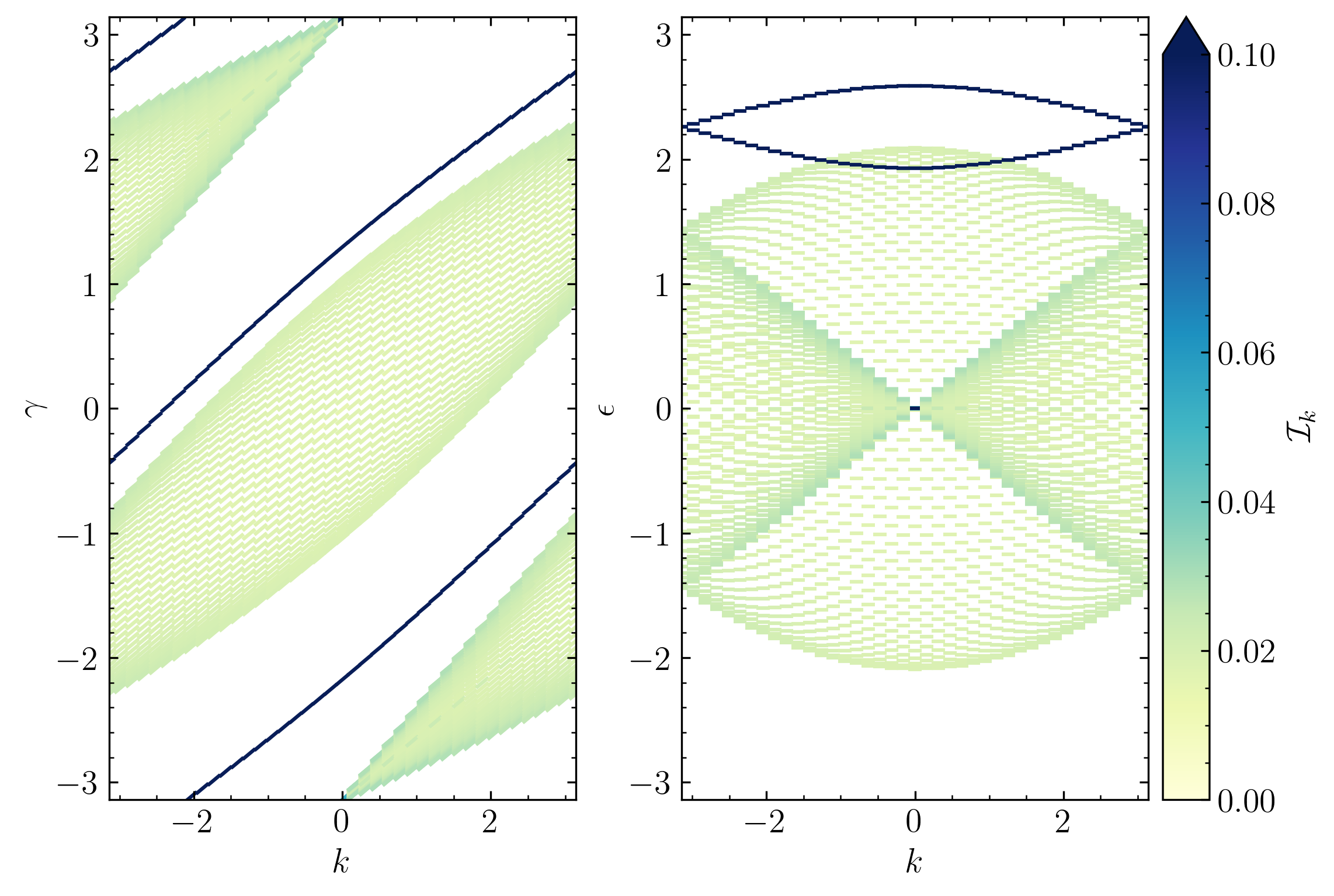}
\caption{Spectrum of the operators ${\mathcal V}_0$ (left) and ${\mathcal U}_0$ (right) in the $N=2$ sector.
The colormap indicates the momentum-resolved inverse participation ratio ${\mathcal I}_k$ (the scale is saturated to a maximum value of $0.10$ for more clear view).
The bands of 2-particle bound states have large ${\mathcal I}_k$ (dark blue), while the 1+1 continuum states have small ${\mathcal I}_k$ (light green).}
\label{fig:chiral2}
\end{figure}

\begin{figure}[h]
\centering
\includegraphics[width=\linewidth]{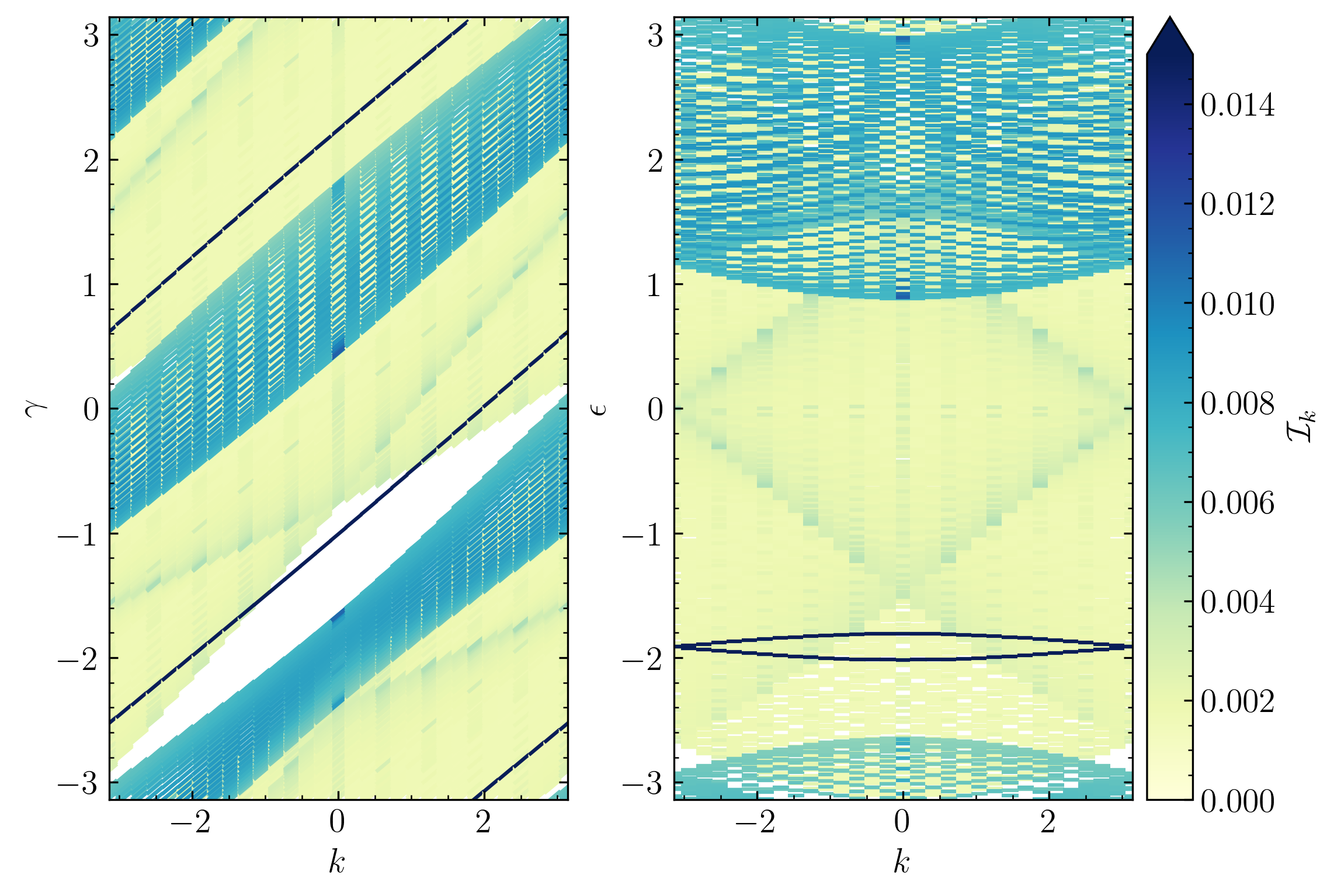}
\caption{Spectrum of the operators $\mathcal V_0$ (left) and $\mathcal U_0$ (right) in the $N=3$ sector.
The colormap indicates the momentum-resolved inverse participation ratio $\mathcal I_k$ (the scale is saturated to a maximum value of $0.15$).
The bands of 3-particle bound states have the largest ${\mathcal I}_k$ (dark blue), followed by the 2+1 continuum (light blue), while the 1+1+1 continuum states have small ${\mathcal I}_k$ (light green).}
\label{fig:chiral3}
\end{figure}

In Figs.~\ref{fig:chiral2} and \ref{fig:chiral3} we plot the spectrum of the operators ${\mathcal V}_0$ and ${\mathcal U}_0$ for the sectors with $N=2$ and $N=3$ particles.
For $N=2$ (Fig.~\ref{fig:chiral2}), we see that the band of bound states is gapped in the spectrum of ${\mathcal V}_0$, while one branch of the band in the spectrum of ${\mathcal U}_0$ is not gapped for $k\approx 0$: the ``folding'' procedure that gives the spectrum of ${\mathcal U}_0$ from the one of ${\mathcal V}_0$ brings part of the bound state band in the continuum of 1+1 states.
However, the presence of a gap in ${\mathcal V}_0$ implies that the bound states are robust to sufficiently small (but finite) perturbations of the Floquet operator that preserve the ``brickwall'' structure and the total number of particles (i.e., to the perturbations of ${\mathcal U}_0$ that correspond to perturbations of the $\mathcal V_0$ operator).
This protection of the bound states is not manifest when one only considers the momentum-resolved spectrum of ${\mathcal U}_0$ in the presence of the overlap with the continuum because such a view does not take into account the non-trivial conserved quantity ${\mathcal V}_0$, namely $[{\mathcal U}_0, {\mathcal V}_0] = 0$; this additional ``symmetry'' is fully taken into account when one considers ${\mathcal V}_0$ together with ${\mathcal T}^2$.

Similar arguments hold for $N=3$: as we see in Fig.~\ref{fig:chiral3}, part of the bound state band is gapped around $k=0$ in the spectrum of ${\mathcal V}_0$.
This part of the band is robust against any (small) perturbation of ${\mathcal V}_0$, despite the absence of a gap in the spectrum of ${\mathcal U}_0$.
Note, however, that the comb teeth perturbation considered in the main text following the GQAI experiments does not preserve the brickwall structure that is crucial for the reduction of the full problem to ${\mathcal V}_0$, and the above protection does not operate in this case.

\section{Detailed understanding of features in the spectrum and the robust bound states in the Hamiltonian system}
\label{sec:detailed}

In this Appendix, we discuss in detail the spectrum of the Hamiltonian system, focusing in particular on the 2-particle and 3-particle bound states for the parameters of interest.
As already mentioned in the main text, except for a small part of the $k$-space in the former case, these are separated from the rest of the spectrum by finite gaps and are hence robust (sharply defined) bound states in the thermodynamic limit.
This allows fully controlled treatment of the bound states and essentially complete understanding of their characters.
While the 3-particle bound states cease to be sharp in the Floquet system of interest, for reasons discussed in the main text the Hamiltonian system with modest folding in the quasienergy space still provides reasonable approximations, allowing much of the intuition about the spectral properties and bound states of the Hamiltonian case to be transferred to the Floquet case.

Hamiltonian spectra shown in this Appendix in Fig.~\ref{fig:Ham_pert1} and top panels of Figs.~\ref{fig:Ham_pert2} and \ref{fig:Ham_pert3} are the same as in the corresponding Hamiltonian panels in Figs.~\ref{fig:spectra12} and \ref{fig:spectrum3}, except that they are not multiplied by $(-\pi/6)$ and are instead presented here in the Hamiltonian energy units.
At the expense of some repetition, this significantly simplifies referencing features in the spectrum and also puts the bound states at the bottom of the spectrum, allowing us to readily use intuition from familiar perturbation theory/effective Hamiltonian tools near ground states.
In quantitative demonstrations below, we use the same parameters $w=1$, $u=u'=4w$, varying $w'$, as in the main text.

For easy reference, we first list the exact dispersions of the 2-particle ($N=2$ sector) and 3-particle ($N=3$ sector) bound states in the unperturbed Hamiltonian $H_0$, Eq.~(\ref{eq:Hint}), translated from the known results for the XXZ chain~\cite{Takahashi1972, Fowler1981, Sutherland_book, Ganahl2012,franchini2017introduction}:
\begin{align}
& \epsilon_{\text{2-string}}(k_\text{chain}) = -u - \frac{2w^2}{u} - \frac{2w^2}{u} \cos(k_\text{chain}),
\label{eq:e2string}\\
& \epsilon_{\text{3-string}}(k_\text{chain}) = -2u - \frac{2uw^2}{u^2 - w^2} - \frac{2w^3}{u^2 - w^2} \cos(k_\text{chain}), 
\label{eq:e3string}
\end{align}
where $k_\text{chain}$ refers to the natural momentum on the chain with translation symmetry by one site \footnote{For  general $\ell$-string, the dispersion relation reads
\begin{equation*}
\epsilon_{\ell\text{-string}}(k_\text{chain}) = 2w \frac{\sinh(\nu)}{\sinh(\ell\nu)}\Big[\cosh(\ell\nu) - \cos(k_\text{chain})\Big] - \ell u,
\end{equation*}
where $\cosh(\nu)=u/(2w)$.}.
For $u=4w$, the 2-string and 3-string bound states are separated from continuum states by finite gaps at each $k_\text{chain}$;
the top of the 2-string bound state band at $k_\text{chain} = \pi$ happens to coincide with the bottom of the 1+1 continuum at $k=0$, while the top of the 3-string bound state band lies significantly below the bottom of the corresponding 2+1 continuum (which lies below the 1+1+1 continuum).

Turning to the perturbed problem with the additional sites (the comb lattice in Fig.~\ref{fig:lattice_and_boundstates}), we start with some general remarks.
The conserved total number of particles in the system is denoted $N$.
When $w'=0$, the particle number on the original chain is separately conserved and is denoted $N_{1\cup 2}$; furthermore, the particle number on each $(R,1')$ site is conserved (we often use a more crude conserved number $N_{1'} = N - N_{1\cup 2}$ to group states).
In this case, in the sector with $N_{1'} = 0$, the $u'$ term does not operate at all and the Hamiltonian is equivalent to the original integrable chain for any $u'$.
Thus, in this sector, the $u'$ part by itself does not break the integrability, and we sometimes refer to the model with $w'=0$ as integrable. 
On the other hand, still keeping $w'=0$, in sectors with $N_{1'} > 0$, the occupations of the $1'$ sites do not fluctuate but create static potentials $(-u) < 0$ for the chain particles on the $1$ sites connected to the occupied $1'$ sites. 
The model for the particles on the chain is non-integrable because of these static ``impurity potentials'', although for small $N_{1'}$ it is possible to use some intuition/results from the integrable model away from the impurities.

\subsection{One particle}
\begin{figure}[h]
\centering
\includegraphics[width=\linewidth]{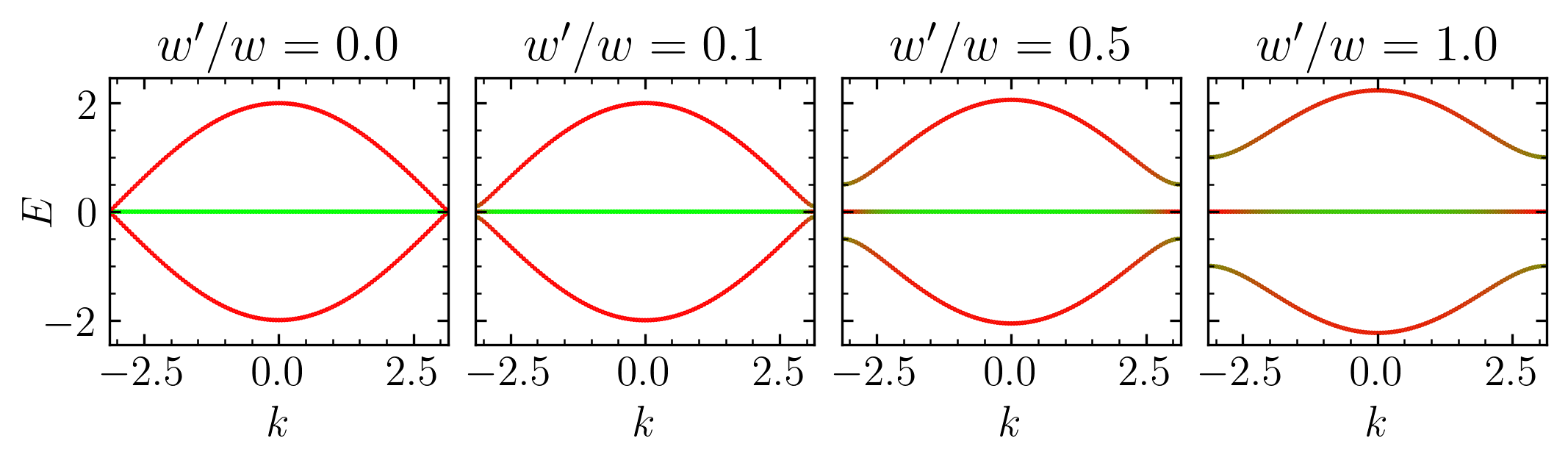}
\caption{Single particle ($N=1$) spectrum of $H$ for $w=1$, $u=u'=4$.
The color mixing represents the weights on the different sectors of the unperturbed model $w'=0$, as described in the caption of Fig.~\ref{fig:spectra12}.}
\label{fig:Ham_pert1}
\end{figure}

In the case of total of one particle in the system, the $u$ and $u'$ terms do not operate at all and the problem reduces to a single-particle problem with hopping amplitude $w$ along the chain and $w'$ on the $1$-$1'$ links connecting to the extra sites.
This problem is easily solved and has three bands 
\begin{equation}
\epsilon_{\pm}(k) = \pm \sqrt{\left(2w \cos\frac{k}{2}\right)^2 + (w')^2} ~, \quad 
\epsilon_0(k) = 0 ~,
\end{equation}
which are shown in Fig.~\ref{fig:Ham_pert1}.
The symmetry $E \leftrightarrow -E$ around the zero energy as well as the origin of the entire $E=0$ band are due to the bipartite hopping nature of this single-particle problem:
There are no on-site potentials and the hoppings connect only sites on different sublattices $A \equiv \{(R,2), (R,1')\}$ and $B \equiv \{(R,1)\}$.
Since there are twice as many $A$ sites as there are $B$ sites, $M_A = 2\Luc, M_B = \Luc$, we expect at least $M_A - M_B = \Luc$ zero-energy states residing entirely on the $A$ sublattice.

\subsection{Two particles}

\begin{figure}[h]
\centering
\includegraphics[width=\linewidth]{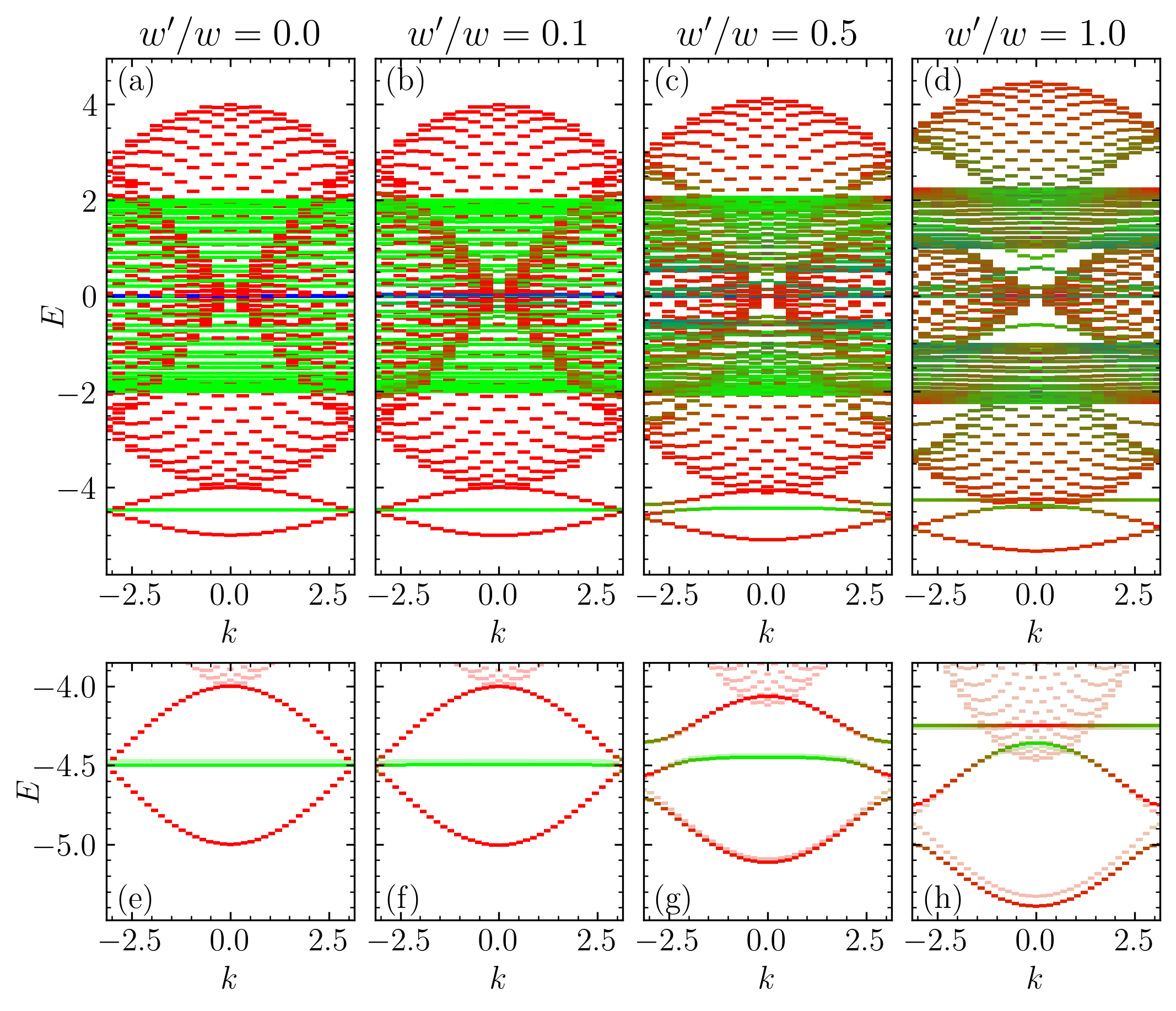}
\caption{(a-d) Two particle ($N=2$) spectrum of $H$ for $w=1$, $u=u'=4$,  $\Luc=20$ .
The color mixing represents the weights on the different sectors of the unperturbed model $w'=0$, as described in the caption of Fig.~\ref{fig:spectra12}.
(e-h) Zoom-out of the region containing the 2-particle bound states.
Semi-transparent lines: 2-particle bound states (and nearby other states) in the spectrum of $H$ for the same parameters as in (a-d) but larger $\Luc=36$ used for smoothness.
Solid lines: Spectrum of the effective Hamiltonian Eq.~(\ref{eq:Heff_Ham2}) derived for $w,w' \ll u, u'$.  The exact spectrum and the results of the effective model have almost perfect overlap for all the three bands of bound states.}
\label{fig:Ham_pert2}
\end{figure}

Figure~\ref{fig:Ham_pert2}(a-d) shows the momentum resolved exact diagonalization results in the Hamiltonian units, with fixed $u = u' = 4w$, where we vary $w'$ from $w' = 0$ to $w' = w$; we set $w = 1$.
We do not have an exact solution for general $w'$ in this case, but we can understand rough features and in particular the 2-particle bound states by developing from certain controlled limits, which we now describe.

\subsubsection{$w'=0$, general $w$}
The eigenstates can be divided into three main groups (sectors) defined by specifying the (conserved) number of particles on the chain, denoted $N_{1\cup 2}$, and on the extra sites, denoted $N_{1'}$:

\underline{$(N_{1\cup 2}, N_{1'}) = (2, 0)$:} There are $\Luc(2\Luc-1)$ such states and they are marked red in panel~(a) in Fig.~\ref{fig:Ham_pert2}.
This sector is equivalent to the unperturbed integrable model on the original chain.
The corresponding states in Fig.~\ref{fig:Ham_pert2}(a) represent simply folding of the original chain spectrum to the new Brillouin zone.
We clearly see the 2-particle continuum spanning energy window $(-4w, 4w)$ in the thermodynamic limit.
We are mainly interested in the lowest-energy states forming two bands of the 2-string bound states. 
In the original chain, the Hamiltonian is invariant under translation by a single site, and the bound states form a single band in the corresponding Brillouin zone, with dispersion given by Eq.~(\ref{eq:e2string}).
This band is here folded to the new Brillouin zone, resulting in two distinct bands.

\underline{$(N_{1\cup 2}, N_{1'}) = (1, 1)$:} There are $2\Luc^2$ such states and they are marked green in panel~(a) in Fig.~\ref{fig:Ham_pert2}.
These can be further subdivided into subgroups labelled by a location $(R_0, 1')$ of the one particle on the $1'$ sites, which remains completely localized since $w'=0$.
The spectra are identical for different $R_0$; each energy level is hence repeated $\Luc$ times in the full spectrum and gives a flat band in the energy vs momentum plot in Fig.~\ref{fig:Ham_pert2}(a).

For a fixed $R_0$, we have a problem of one particle hopping on the original chain of $2\Luc$ sites with the hopping amplitude $w$ and a single attractive ``impurity potential'' $(-u') < 0$ felt by the particle  when it is on the site $(R_0,1)$.
In this hopping problem, we expect one localized state near the impurity potential and $2\Luc - 1$ delocalized states.
The delocalized states span energy window of $(-2w,2w)$ in the thermodynamic limit.

One can easily solve for the localized state $\psi_{\text{localized}}(j) = C e^{-\kappa |j - j_0|}$ in the thermodynamic limit, where $j$ labels sites as in the original chain and $j_0$ is the corresponding label of the site $(R_0,1)$.
The localized state energy and the rate of the wavefunction decay per lattice site are
\begin{align}
& \epsilon_{\text{localized}} = -\sqrt{(u')^2 + 4w^2} ~, \\ 
& e^{-\kappa} = \frac{\sqrt{(u')^2 + 4w^2} - u'}{2w} ~.
\end{align}
In the full system, this can be viewed as a 2-particle bound state with one of the particles immobile on the $(R_0,1')$ site and the other residing on the chain sites but dynamically bound to the immobile particle.
For our numerical parameters $u'=4w$ we obtain $\epsilon_{\text{localized}} = -2\sqrt{5}w \approx -4.472$ and $e^{-\kappa} = \sqrt{5} - 2 \approx 0.236$; thus, the localization length is $0.693$ of the original chain lattice spacing, which means we have a fairly compact bound state, and the above expression for $\epsilon_{\text{localized}}$ is very accurate even for relatively small sizes.

As mentioned earlier, this analysis gives identical spectra for each of the $\Luc$ possible locations $R_0$, which results in $\Luc$-fold degeneracy for each found eigenvalue.
Each such eigenvalue gives rise to a completely flat band when the full spectrum is resolved in momentum.
Looking at the green states in Fig.~\ref{fig:Ham_pert2}(a) marking the present sector, we see the corresponding dense set of flat bands in the energy window $(-2w,2w)$ for the delocalized states and the flat band near $\epsilon_{\text{localized}}$ for the 2-particle bound states.
It is a numerical accident for the specific parameters that this energy is very close to where the two red 2-particle bound state bands meet, which happens at $ \epsilon_{\text{2-string}}(k_\text{chain}=\pi/2) = -u - 2w^2/u = -4.5$ [using Eq.~(\ref{eq:e2string})].
Note that in our case where the dominant interaction binding energies are taken to be the same, $u' = u$ (motivated by the GQAI experiments), we expect all 2-particle bound states (i.e., chain-chain and chain-extra site) to be roughly in the same ballpark, while the precise band locations depend on further dynamical details from the hopping energy.

\underline{$(N_{1\cup 2}, N_{1'}) = (0, 2)$:} There are $\Luc(\Luc-1)/2$ such states where both particles are on the $1'$ sites.
These states all have energy $0$ and are marked blue in panel~(a) in Fig.~\ref{fig:Ham_pert2}.
They are not important in our considerations below focusing on the bottom of the spectrum near where the 2-string bound states reside.

\subsubsection{Small $w'$, general $w$: qualitative considerations}
Now we add non-zero but small $w'$.
The $w'$ term does not act within the above sectors but connects the $(N_{1\cup 2}, N_{1'}) = (2,0)$ and $(1,1)$ sectors [and also the $(1,1)$ and $(0,2)$ sectors].
This roughly explains why in Fig.~\ref{fig:Ham_pert2}~(a-d) outside of the $(-2w, 2w)$ energy window the $(2,0)$ sector states are not strongly affected for $w'=0.1$ and $w'=0.5$ (even all the way to $w'=1$), except close to the boundary of this window and to the bound states from the $(N_{1\cup 2}, N_{1^\prime}) = (1, 1)$ sector. 
The latter bound states (which we refer to as chain-extra site bound states) couple with the bound states from the $(N_{1\cup 2}, N_{1^\prime}) = (2, 0)$ sector (chain-chain bound states) and with the bottom of the two-particle continuum from the $(N_{1\cup 2}, N_{1^\prime}) = (2, 0)$ sector.
There are some additional features in the middle of the spectrum that arise with increasing $w'$, but these are not of direct interest to us and are not studied further.

From now on we focus on the 2-string bound states.
One of the effects of adding $w'$ on the chain-chain bound states
is that they can lower their energy by virtual processes where one of the particles hops off the chain onto a $1'$ site and back.
This lowering of the energy is roughly $w'^2/u$ and is visible in Fig.~\ref{fig:Ham_pert2}~(b-d) for the red 2-string bound state bands.
On the other hand, the chain-extra site bound states, in the limit of very compact bound states, do not have this mechanism and to this order their energy would remain unchanged; such tendency is visible in Fig.~\ref{fig:Ham_pert2}~(b-d) for the green 2-string bound states.
However, the chain-extra site bound states can hybridize with the chain-chain bound states with amplitude $O(ww'/u)$ particularly when their energies are close, which happens near the wavevector $\pi$ at $w'=0$ and moves to smaller wavevectors as $w'$ increases and the chain-chain bound states move to lower energies.
This hybridization with the moving central location is visible in the progression in Fig.~\ref{fig:Ham_pert2}~(a-d) as we increase $w'$, while for $w'=1$ the descendants of the chain-extra site bound states no longer overlap with the descendants of the chain-chain bound states.

From the figure, we see that the predominantly green 2-string bound states survive for momenta sufficiently away from zero even for $w'=1$, while they are in the continuum of states for momenta close to zero and will not survive in the thermodynamic limit.
For the predominantly red 2-string bound states, only states in the upper band with momenta close to zero enter the continuum spectrum and will not survive in the thermodynamic limit (although their decay rate is likely very small), while the rest of this upper band and all of the lower band 2-string bound states clearly survive in the thermodynamic limit protected from the continuum by gaps at the corresponding momenta.

\subsubsection{Perturbative treatment for $w, w' \ll u, u'$}
Some of the above qualitative arguments can be made more precise by taking the limit $w, w' \ll u, u'$, which we discuss here for completeness.
In the absence of the hoppings, the lowest-energy states are the following dimer states specified by the particle locations on the full system (chain + extra sites), and depicted in Fig.~\ref{fig:lattice_and_boundstates}(c):
\begin{align}
& D_{\text{I}}(R) = [(R,1), (R,2)] ~, \\
& D_{\text{II}}(R) = [(R,1), (R-1,2)] ~,
\qquad \epsilon_{\text{I,II}} = -u ~; \\
& D_{\text{III}}(R) = [(R,1'), (R,1)] ~, \qquad \epsilon_{\text{III}} = -u' ~.
\end{align}

In the second-order perturbation theory, first we have diagonal corrections, which for later convenience we write in the ket-bra notation for the dimer states associated with location $R$ as defined above:
\begin{align*}
\hat{h}^{\text{eff}}_{\text{diag}}(R) =& -\frac{2w^2 + w'^2}{u} \sum_{\ell=\text{I,II}}\ketbra{D_{\ell}(R)} \\
& -\frac{2w^2}{u'} \ketbra{D_{\text{III}}(R)} ~.
\end{align*}
Here we explicitly see the claimed lowering of the energies of the chain-chain dimers $D_{\text{I}}, D_{\text{II}}$ via virtual fluctuations involving $w'$ hops, while no such lowering is present for the chain-extra site dimers $D_{\text{III}}$.

Next, in the same order, the chain-chain dimers can hop along the chain with amplitude $w^2/u$.
Explicitly, in the above notation,
\begin{align*}
\hat{H}^{\text{eff}}_{\text{hop}} = -\frac{w^2}{u} \sum_R \big[ & \ketbra{D_{\text{II}}(R)}{D_{\text{I}}(R)} + \text{H.c.} \\
& + \ketbra{D_{\text{I}}(R)}{D_{\text{II}}(R+1)} + \text{H.c.} \big] ~.
\end{align*}
Note that the chain-extra site dimers cannot hop by themselves at this order [for $u \neq u'$, such hopping can appear only at $O(w'^2 w^4/u'^5$)].

Finally, for \underline{$u'=u$}, assumed from now on, where we need to do degenerate perturbation theory involving all $D_{\text{I}}(R)$, $D_{\text{II}}(R)$, $D_{\text{III}}(R)$, the $D_{\text{III}}(R)$ can convert to $D_{\text{I}}(R)$ or $D_{\text{II}}(R)$ and vice versa:
\begin{align*}
\hat{h}^{\text{eff}}_{\text{I,II--III}}(R) = -\frac{ww'}{u} \big[&\ketbra{D_{\text{I}}(R)}{D_{\text{III}}(R)} + \text{H.c.} \\ 
&+ \ketbra{D_{\text{II}}(R)}{D_{\text{III}}(R)} + \text{H.c.} \big] ~.
\end{align*}

Putting everything together, we have an effective Hamiltonian 
\begin{equation}
H^{\text{eff}} = \hat{H}^{\text{eff}}_{\text{hop}} + \sum_R \left[\hat{h}^{\text{eff}}_{\text{diag}}(R) + \hat{h}^{\text{eff}}_{\text{I,II--III}}(R) \right] ~.
\label{eq:Heff_Ham2}
\end{equation}
By going to momentum space, we obtain a $3 \times 3$ matrix that is easy to diagonalize numerically and explore the evolution of the 3 bands, e.g., as one varies $w'$ relative to $w$.
The results are shown in Fig.~\ref{fig:Ham_pert2}(e-h) and capture roughly the behavior seen for the full problem and discussed qualitatively earlier:
As $w'/w$ increases from $0$ to $1$, the chain-chain dimer bands move to lower energies while at the same time particularly the higher-energy one of them hybridizes significantly with the chain-extra site dimer band, as seen in the mixing of the colors of these bands in Fig.~\ref{fig:Ham_pert2}.
This model does not capture all detailed features seen in the figure, e.g., the apparent very small gaps between the two lowest bands at $k=\pm \pi$, presumably because of an inaccurate treatment of the sizable $w$ used in the figure; however, interestingly, at $w'=w$ it gives a completely flat topmost band capturing the nearly flat uppermost 2-string bound state band in the figure.
We do not  consider further details here since in any case this model does not include the eventual (small) overlaps with the two-particle continuum, where we have to use the full problem ED results.

\subsection{Three particles}
\begin{figure}
\centering
\includegraphics[width=\linewidth]{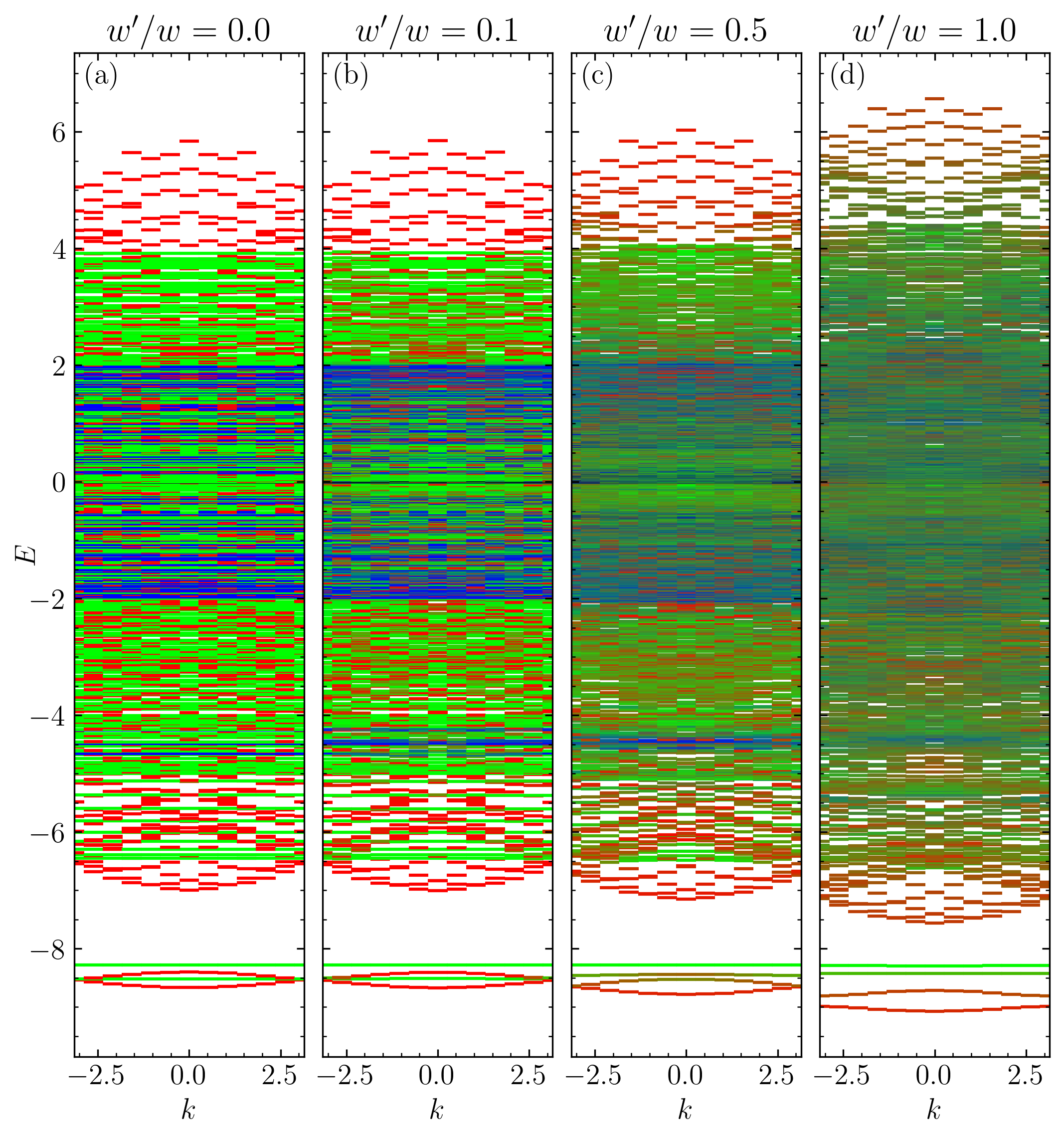}
\includegraphics[width=\linewidth]{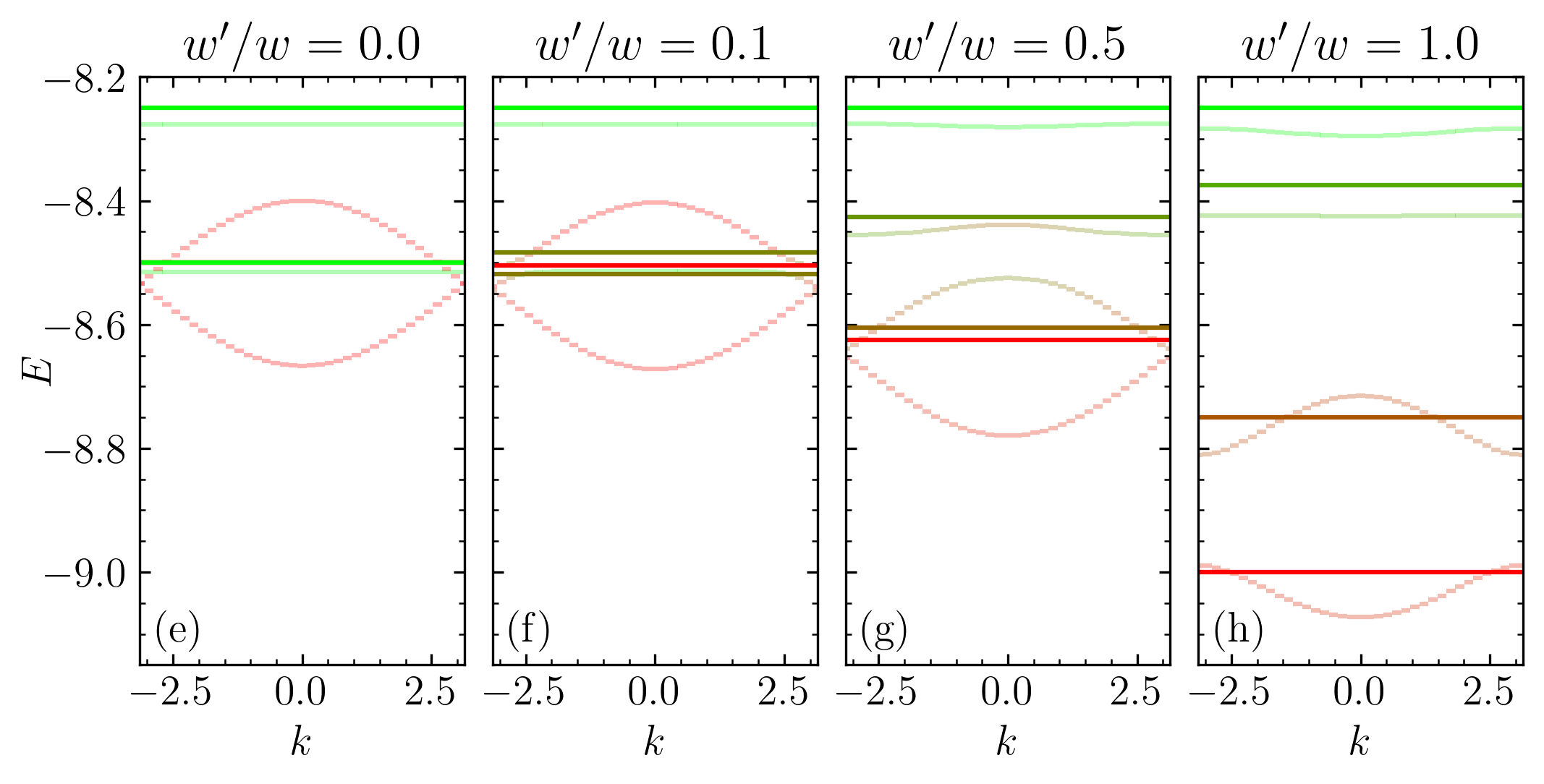}
\caption{(a-d) Three particle ($N=3$) spectrum of $H$ for $w=1$, $u=u'=4$, $\Luc=12$. The color mixing represents the weights on the different sectors of the unperturbed model $w'=0$, as described in the caption of Fig.~\ref{fig:spectrum3}. 
(e-h) Zoom-out of the region containing the 3-particle bound states.
Semi-transparent lines: bands of the 3-particle bound states for the same parameters as in (a-d) but larger $\Luc=36$. 
Solid lines: (flat) bands of the 3-particle bound states from solving the second-order effective Hamiltonian, Eqs.~(\ref{eq:epsilonI_Ham3})-(\ref{eq:epsilonIIs_Ham3}).
In panel (e), the lowest flat band of the effective Hamiltonian ($E=-2u-2w^2/u=-8.5$) is three-fold degenerate:  one band belongs to the $(N_{1\cup 2}, N_{1'}) = (2, 1)$ sector and the others are in the $(N_{1\cup 2}, N_{1'}) = (3, 0)$ sector (only the first one is visible, in green).}
\label{fig:Ham_pert3}
\end{figure}

\subsubsection{$w'=0$, general $w$}
The eigenstates can be divided into four main groups defined by specifying the number of particles on the chain  $N_{1\cup 2}$ and on the extra sites $N_{1'}$:

\underline{$(N_{1\cup 2}, N_{1'}) = (3, 0)$}: There are $\Luc(2\Luc-1)(2\Luc-2)/3$ such states and they are marked red in panel~(a) in Fig.~\ref{fig:Ham_pert3}.
This sector is equivalent to the unperturbed integrable model on the original chain.
The corresponding states in Fig.~\ref{fig:Ham_pert3}(a) represent simply folding of the earlier chain spectrum to the new Brilloin zone.
We are primarily interested in the lowest-energy states forming two bands of the 3-string bound states representing the single band of these in the original chain, with dispersion given by Eq.~(\ref{eq:e3string}), folded to the new Brillouin zone.

\underline{$(N_{1\cup 2}, N_{1'}) = (2, 1)$}: There are $\Luc^2(2\Luc-1)$ such states and they are marked green in panel~(a) in Fig.~\ref{fig:Ham_pert3}.
These can be further subdivided into subgroups labelled by a location $(R_0, 1')$ of the one particle on the $1'$ sites, which remains completely localized under such Hamiltonian.
The spectra are identical for different $R_0$; each energy is hence repeated $\Luc$ times in the full spectrum and gives a flat band in the energy vs momentum plot in Fig.~\ref{fig:Ham_pert3}(a).

For a fixed $R_0$, we obtain a problem with two particles on the original chain but with an attractive ``impurity potential'' $(-u') < 0$ on one site $j_0$ corresponding to $(R_0,1)$.
Away from the impurity we have the 2-particle continuum (covering energy window of $[-4w, 4w] = [-4,4]$ for $w = 1$) as well as the band of 2-string bound states (energy window of $[-u-4w^2/u, -u] = [-4,-5]$ for $u = 4w$ used here).
The attractive impurity will lead to appearance of some localized states out of these.

We are mainly interested in the effect of the impurity on the 2-string bound states.
We can roughly model these as dimers [covering sites $(j,j+1)$] hopping on the original chain [hops $(j,j+1) \leftrightarrow (j+1,j+2)$] with amplitude $w_{\text{dimer}}$ and background energy $\bar{\epsilon}_{\text{dimer}}$, which we can estimate by fitting the exact dispersion of the 2-string bound states in the integrable model, Eq.~(\ref{eq:e2string}), as
\begin{align}
& \epsilon_{\text{2-string}}(k_\text{chain}) = \bar{\epsilon}_{\text{dimer}} - 2 w_{\text{dimer}} \cos(k_\text{chain}) ~, \\
& \bar{\epsilon}_{\text{dimer}} = -u - \frac{2w^2}{u}~, \qquad w_{\text{dimer}} = \frac{w^2}{u} ~.
\end{align}
The dimer feels the attractive impurity at $j_0$ for two of its positions, $(j_0-1,j_0)$ and $(j_0,j_0+1)$, and the problem is mathematically equivalent to a point particle hopping on a lattice with potential $(-u')$ on two neighboring sites.
On an infinite lattice and in this model of a rigid dimer, we can solve analytically for exponentially localized states and obtain energies
\begin{align*}
& \epsilon_{\text{loc.dimer, sym.}} = \bar{\epsilon}_{\text{dimer}} - u' - w_{\text{dimer}} - \frac{w_{\text{dimer}}^2}{u' + w_{\text{dimer}}} ~, \\
& \epsilon_{\text{loc.dimer, anti-sym.}} = \bar{\epsilon}_{\text{dimer}} - u' + w_{\text{dimer}} - \frac{w_{\text{dimer}}^2}{u' - w_{\text{dimer}}} ~.
\end{align*}
The first localized state is always present and is symmetric around $j_0$, while the second localized state is present if $u' > 2w_{\text{dimer}}$ (which is satisfied in our problem) and is anti-symmetric around $j_0$.
These localized dimer states can be viewed as 3-particle bound states where one of the particles resides on the extra sites; we will often refer to these also as 3-string bound states.
The two lowest green flat bands in Fig.~\ref{fig:Ham_pert3}(a) [close to the red weakly dispersive bands of the 3-string bound states from the $(N_{1\cup 2}, N_{1^\prime}) = (3, 0)$ sector] correspond to these states viewed as bands once we include all different $R_0$.
Estimating $w_{\text{dimer}} = 0.25$ from the 2-string dispersion, we can estimate the splitting of about $2w_{\text{dimer}} \approx 0.5$, which is somewhat larger than the actual splitting of $\approx 0.25$ between the corresponding green bands in the figure; the inaccuracy is likely due to crude modelling of the 2-string bound state by the above dimer picture.

We will not consider any other localized states in this group, which will be below the delocalized continuum of states but significantly above the 3-particle bound states of interest.
Of interest for connecting with the Floquet case are the highest-energy states in this group, which are near the energy $\approx 4$:
If we take the Hamiltonian spectrum and multiply it by ``time'' $\pi/6$ as a rough estimate to connect with the GQAI Floquet experiment as done in the main text, the 3-string bound states from the sector $(N_{1\cup 2}, N_{1^\prime}) = (3, 0)$, upon ``folding'' modulo $2\pi$ in the Floquet quasienergy space, would land among the states that are in the window $\approx [3,4]$ in the Hamiltonian spectrum in this figure.
The nature of these states from the $(N_{1\cup 2}, N_{1^\prime}) = (2, 1)$ sector is as follows: there is one particle on $1'$ and two particles on the chain in scattering states staying away from the $1'$ particle and from each other.
We expect that matrix elements of the $w'$ perturbation between these and the 3-string bound states from the $(N_{1\cup 2}, N_{1^\prime}) = (3, 0)$ sector are very small, since the $w'$ term would break such a 3-string state into a particle on a $1'$ site and a nearby dimer, and both the proximity to the ``impurity'' and the proximity of the two particles on the chain has low probability in the described scattering states. 
In Sec.~\ref{sec:perturbative} we use such characters of the states involved to obtain scaling of these matrix elements with $\Luc$.
These matrix elements are not relevant at all in the Hamiltonian problem where the 3-string bound states and these states are separated by a large energy difference, but they are important for understanding robustness of the 3-string bound states in the Floquet problem.
As discussed further in Sec.~\ref{sec:perturbative}, the density of states is also an important factor in estimating the decay rate of the bound states, requiring a quantitative study as presented in the main text.

\underline{$(N_{1\cup 2}, N_{1'}) = (1, 2)$}: There are $\Luc^2(\Luc-1)$ such states and they are marked blue in panel~(a) in Fig.~\ref{fig:Ham_pert3}.
Here two particles are immobile on some $1'$ sites, say $(R_0,1')$ and $(\tilde{R}_0,1')$, while the remaining particle is moving on the chain and sees the immobile particles as impurity potentials $(-u')$ at $j_0$ and $\tilde{j}_0$ corresponding to $(R_0,1)$ and $(\tilde{R}_0,1)$.
Away from the impurities we have free propagation with the band covering energy window $[-2w,2w]$, while the impurities will localize some states out of this band.
Since $u'$ is sufficiently large, we expect that the two impurities will lead to two localized states out of the band, whose energies will depend somewhat on specific relative position of the impurities but will be independent of the overall shift of $R_0$ and $\tilde{R}_0$, leading to flat bands.
These localized-state energies are visible in the spectrum around energy $\approx -4.5$.
Neither of the blue states are important for understanding the 3-string bound states in the Hamiltonian and the Floquet cases.

\underline{$(N_{1\cup 2}, N_{1'}) = (0, 3)$}: There are $\Luc(\Luc-1)(\Luc-2)/6$ such states and they are marked black in panel (a) in Fig.~\ref{fig:Ham_pert3}.
These have zero energy and are not of much interest for the study of the 3-string bound states.

\subsubsection{Small $w'$, general $w$: qualitative considerations}
We now focus solely on the 3-string bound states.
One of the important effects of adding $w'$ seen in Fig.~\ref{fig:Ham_pert3} is that the energies of the 3-string states whose particles reside on the chain go down with increasing $w'$, while the energies of the 3-string states that have one particle on the extra sites remain essentially unchanged.
We can understand this simply as follows.
For the former 3-string states, which we will refer to as chain 3-string states [see also pictures $T_{\text{I}}$ and $T_{\text{II}}$ in Fig.~\ref{fig:lattice_and_boundstates}(d) in the tight trimer limit], the non-zero $w'$ allows virtual fluctuations involving hopping of one of the particles off the chain onto a nearby extra site site and back, leading to lowering of the energy.
On the other hand, for the latter 3-string states, which we will refer to  as dimer-$1'$ 3-string bound states [see also pictures $T_{\text{III}}$ and $T_{\text{IV}}$ in Fig.~\ref{fig:lattice_and_boundstates}(d)], such virtual fluctuations are not available when the dimer is tightly bound to the particle on the $1'$ site.
The antisymmetric dimer-$1'$ 3-string bound state [schematically, antisymmetric combination of $T_{\text{III}}$ and $T_{\text{IV}}$, see next subsection for more details] is separated from the chain 3-string bound states already at $w'=0$, and the separation only increases with adding $w'$.
From the evolution in panels~(a)-(d) in Fig.~\ref{fig:Ham_pert3} and the essentially unchanged bright green color of the corresponding band even at $w'=w$, we conclude that the character of this state remains essentially unchanged.
On the other hand, the symmetric dimer-$1'$ 3-string bound state energy at $w'=0$ is close to the lowest energy of the upper chain 3-string band near $k = \pm \pi$, and as the latter moves down upon increasing $w'$ the two bands overlap and mix particularly near momenta where their energies are close.
By the time $w'$ reaches value of $0.5$ the two bands are already separated and stay separated afterwards, also pushing a bit away from each other by level repulsion.
We can then conclude that the darker-green band is roughly the symmetric dimer-$1'$ 3-string bound state with small admixture of fluctuations to the chain 3-string bound state.
Both green bands remain essentially flat since moving such a dimer-$1'$ 3-string bound state requires at least four $w$-hops and two $w'$-hops, i.e., high order perturbation theory in the hoppings relative to the interactions.

Finally, when the chain 3-string bound states are well separated from the dimer-$1'$ 3-string bound states, we can understand the effect of $w'$ on the former in more detail as follows:
We start with the picture of a trimer hopping on the chain.
In the presence of the extra sites, the trimer has two inequivalent positions: one where both ends of the trimer are over extra sites and the other where the middle of the trimer is over an extra site.
In the former case, virtual fluctuations lower the energy of the trimer by $2 w'^2/u$ while in the latter case they lower the energy by only $w'^2/u$.
Thus, we can model the effect of small $w'$ on the trimer as a potential $-3w'^2/(2u) + (-1)^j w'^2/(2u)$.
This has both a uniform part shifting everything down in energy and a staggered part that will open a gap of roughly $w'^2/u$ at $\pm \pi/2$ in the original chain Brilloin zone.
Folded to the new Brilloin zone, we have a picture roughly similar to the two red bands with the gap near the corresponding Brilloin zone boundaries.
The above picture gives an estimate of the gap between the bands as $w'^2/u = 0.25$ at $w'=w=1$, $u=4$, which is somewhat larger but is still fairly close to the observed gap in Fig.~\ref{fig:Ham_pert3}(d).
The inaccuracy is likely due to approximations when modelling the 3-string states by rigid trimers and also due to a larger admixture of the dimer-$1'$ in the upper band, as seen in the difference of the colors between the upper and lower red bands, which is not treated in the above trimer model.

To summarize, in this Hamiltonian system, besides the exact numerical results showing stable 3-string bound states well-separated from the continuum, we also have a fairly complete qualitative picture of the 3-string bound states.
Note that even though we expect no decay of the 3-string bound states at $w'=1$ similar to the integrable $w'=0$ case, there are still significant quantitative differences in the 3-string propagation properties.
For example, the maximal band velocity that determines the propagation wavefront is significantly smaller in the perturbed case.
We expect similar quantitative effects also for (approximate) 3-particle bound states in the Floquet system, which could be tested in experiments.

\subsubsection{Perturbative treatment for $w,w' \ll u,u'$}
Some of the intuition including $w'$ perturbatively relied on the pictures of tightly bound dimers or trimers.
This is formally valid in the regime of small $w \ll u$, and for $w,w' \ll u,u'$ we can perform the corresponding calculations systematically, which we do here for completeness.
In the absence of the hoppings, the lowest energy states are the following trimer states specified by the particle locations on the chain + extra sites system [see Fig.~\ref{fig:lattice_and_boundstates}(d)]:
\begin{align*}
& T_{\text{I}} = [(R,1), (R,2), (R+1,1)] ~, \\
& T_{\text{II}} = [(R-1,2), (R,1), (R,2)] ~, \quad 
\epsilon_{\text{I,II}}^{(0)} = -2u ~; \\
& T_{\text{III}} = [(R,1'), (R,1), (R,2)] ~, \\
& T_{\text{IV}} = [(R-1,2), (R,1), (R,1')] ~, \quad 
\epsilon_{\text{III,IV}}^{(0)} = -u - u' ~.
\end{align*}
The states $T_{\text{I}}, T_{\text{II}}$ form a degenerate manifold, and so do $T_{\text{III}}, T_{\text{IV}}$.
The corresponding two manifolds are separated in energy if $u' \neq u$ and would be treated separately in this case.
On the other hand, they are degenerate if $u' = u$ and should be treated together in this case.

Adding small $w$ and $w'$, the first perturbative corrections appear in quadratic order.
First, there are diagonal corrections appearing from virtual process processes where one of the particles hops away from the other two and then comes back, obtained by simply examining available such virtual moves:
\begin{align*}
& h^{\text{eff}}_{\text{I,I}} = -\frac{2w^2}{u} - \frac{2w'^2}{u} ~, \qquad
h^{\text{eff}}_{\text{II,II}} = -\frac{2w^2}{u} - \frac{w'^2}{2u} ~, \\
& h^{\text{eff}}_{\text{III,III}} = h^{\text{eff}}_{\text{IV,IV}} = -\frac{w^2}{u} - \frac{w^2}{u+u'} ~.
\end{align*}

Next, at this order the above $T_{\text{III}}$ and $T_{\text{IV}}$ at the same $R_0$ get connected with matrix elements
\begin{equation*}
h^{\text{eff}}_{\text{III,IV}} = h^{\text{eff}}_{\text{IV,III}} = -\frac{w^2}{u+u'} ~.   
\end{equation*}
The $T_{\text{III}}$-$T_{\text{IV}}$ block is diagonalized by considering symmetric and anti-symmetric combinations, obtaining
\begin{equation*}
h^{\text{eff}}_{\text{s,s}} = -\frac{w^2}{u} - 2\frac{w^2}{u+u'} ~, \qquad 
h^{\text{eff}}_{\text{a,a}} = -\frac{w^2}{u} ~.
\end{equation*}

Finally, when \underline{$u'=u$}, which we assume from now on, we need to also consider connections betweem $T_{\text{II}}$ and $T_{\text{III}}$, $T_{\text{IV}}$, which appear at this order:
\begin{align*}
& h^{\text{eff}}_{\text{II,III}} = 
h^{\text{eff}}_{\text{III,II}} = h^{\text{eff}}_{\text{II,IV}} =  h^{\text{eff}}_{\text{IV,II}} = 
-\frac{ww'}{2u} \\
& \implies \quad h^{\text{eff}}_{\text{II,s}} = h^{\text{eff}}_{\text{s,II}} = -\frac{ww'}{\sqrt{2} u} ~, \quad
h^{\text{eff}}_{\text{II,a}} = h^{\text{eff}}_{\text{a,II}} = 0 ~.
\end{align*}

We see that at this order the $T_{\text{I}}$ is not coupled with the rest of the states and has the energy,
\begin{equation}
\epsilon_{\text{I}} = -2u - \frac{2w^2}{u} - \frac{2w'^2}{u} ~.
\label{eq:epsilonI_Ham3}
\end{equation}
Next, the anti-symmetric combination of $T_{\text{III}}$ and $T_{\text{IV}}$ also decouples and has the energy, 
\begin{equation}
\epsilon_{\text{a}} = -2u - \frac{w^2}{u} ~.
\label{eq:epsilona_Ham3}
\end{equation}
Finally, the states $T_{\text{II}}$ and the symmetric combination of $T_{\text{III}}$ and $T_{\text{IV}}$ hybridize, producing energies
\begin{equation}
\epsilon_{\text{II-s},\pm} = -2u - \frac{2w^2}{u} - \frac{w'^2}{4u} \pm \sqrt{\left(\frac{w'^2}{4u} \right)^2 + \left(\frac{ww'}{\sqrt{2}u} \right)^2} ~.
\label{eq:epsilonIIs_Ham3}
\end{equation}
The relative location of the energies of these states is in agreement with the preceding qualitative treatment utilizing the integrable model results, modelling 2-string and 3-string bound states as dimers and trimers and the relevant $u'$ interactions as impurity potentials, and adding $w'$ perturbatively.
On the other hand, the results in this subsection are completely systematic for small $w$ and $w'$, and the character of the states becomes particularly simple in this regime.

At this order, the trimer states are immobile, and the corresponding energies $\epsilon_{\text{I}}$, $\epsilon_{\text{a}}$, and $\epsilon_{\text{II-s},\pm}$ are shown as flat bands in Fig.~\ref{fig:Ham_pert3}(e-h) for varying $w'$, capturing rather well the overall locations of the exact 3-particle bound states from ED.
At the next order (cubic in $w,w'$) the trimer states residing entirely on the chain start hopping with amplitude $w^3/u^2$ [compare with Eq.~(\ref{eq:e3string}) expanded to this order], making the corresponding bands dispersive.
On the other hand, the trimers with a particle on the extra sites are not mobile by themselves at this order, but other processes become available, and we will not  concern to derive the full effective Hamiltonian for the bound states and treat the arising dynamics systematically.

\section{Overview of the Floquet spectrum at $\theta'=0$}
\label{app:overview}
\begin{figure}[h]
\centering
\includegraphics[width=\linewidth]{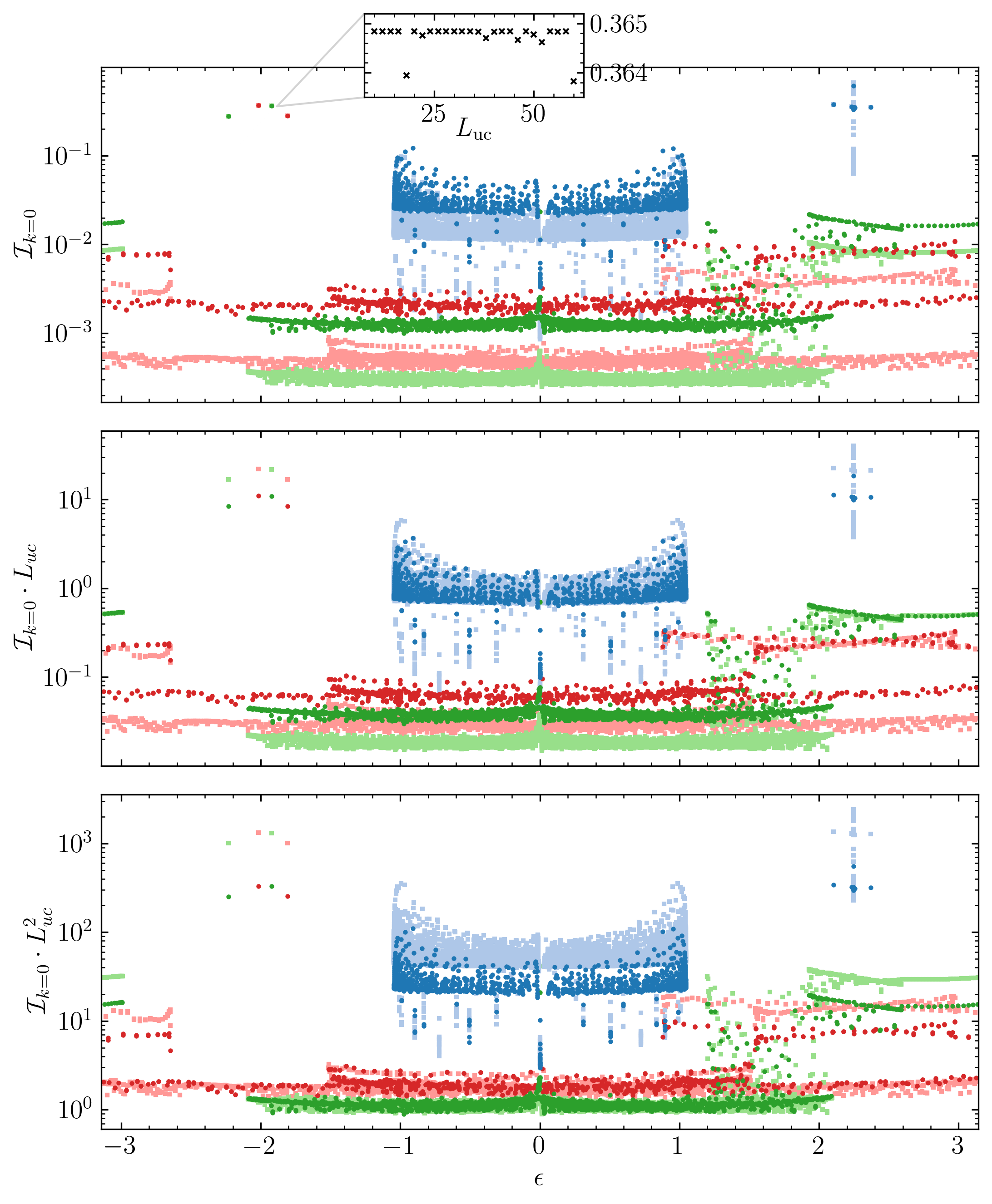}
\caption{Inverse participation ratio in the $k=0$ sector for Floquet eigenstates with quasienergy $\epsilon \in [-\pi,\pi)$, with parameters $\phi=\phi'=2\pi/3$, $\theta=\pi/6$,  $\theta'=0$.
The three panels show the same data multiplied with different powers of $\Luc$.
The red/green/blue colors are used for the sectors $(N_{1\cup 2}, N_{1^\prime}) = (3, 0)$, $(2, 1)$, and $(1, 2)$ respectively.
The light/dark colors are for the different system sizes $\Luc=30$ and $\Luc=60$.
Inset: ${\mathcal I}_{k=0}$ of the bound state with $\epsilon \approx -1.92$ as a function of the system size $\Luc$.
The values for the other bound states have much smaller fluctuations with $\Luc$.}
\label{fig:IPR_resolved}
\end{figure}

In Fig.~\ref{fig:IPR_resolved} we plot the inverse participation ${\mathcal I}_{k=0}$ in the $k=0$ sector of the Floquet eigenstates of the model with $\phi=\phi'=2\pi/3$, $\theta=\pi/6$, $\theta'=0$.
This gives us a finer characterization (and for larger sizes) of the $k=0$ states such as in the $\theta'/\theta=0$ panel in Fig.~\ref{fig:spectra12}, and a wider and more quantitative view of the continuum states than in Fig.~\ref{fig:ipr_spectrum}, which is useful for the discussion of the perturbation theory around this point in the main text.

Plotting ${\mathcal I}_{k=0}$, ${\mathcal I}_{k=0}\cdot \Luc$, and ${\mathcal I}_{k=0}\cdot \Luc^2$ for two different values of the system size, we can easily recognize the different characters of the Floquet eigenstates: (i) 3-particle bound states, which have the same ${\mathcal I}_{k=0}$ irrespective of $\Luc$; (ii) bands of 2+1 continuum states, having ${\mathcal I}_{k=0}\propto \Luc^{-1}$; (iii) bands of 1+1+1 continuum states, with ${\mathcal I}_{k=0}\propto \Luc^{-2}$.

\subsection{Bound states}
In Fig.~\ref{fig:IPR_resolved} we see four bound states with quasienergies $\epsilon \approx -2.23,-2.02, -1.92,-1.81$.
Note that these are the same bound states studied in  Sec.~\ref{sec:perturbative} and labelled $n=0,1,2,3$ there (the values of quasienergies reported there differ from these  by an overall $2\pi$ shift). 
Two of them (in red) belong to the sector with $(N_{1\cup 2}, N_{1^\prime})=(3,0)$ and are protected by the integrability.
The other two (in green) are in the sector with $(N_{1\cup 2}, N_{1^\prime})=(2,1)$: 
In this case, since the particle on the $1'$ site cannot hop, it acts as an impurity. 
These two states then correspond to 2-particle bound states on the chain localized in the potential of the impurity, thus effectively giving 3-particle bound states.
We remark that the model is not integrable in this sector because of the impurity, so the bound states are not protected by simply appealing to the integrability \cite{Zhang_BoundStatesHubbard,sugimoto2023many}.
While the bound state with $\epsilon \approx -2.23$ is protected by a gap (with respect to states in the same sector), the one with $\epsilon \approx -1.92$ lies inside of the 1+1+1 continuum.
Its stability cannot be attributed to any simple mechanisms, but similar studies of integrable models with an impurity have shown the existence of bound states in the continuum, even if the impurity breaks the integrability.
We leave the question of the stability of this bound state for future work.
We note, however, that the inverse participation ratio computed for different system sizes shows much larger fluctuations for this state than for the other bound states (see inset in Fig.~\ref{fig:IPR_resolved}).
These fluctuations may be attributed to a weak hybridization with states in the continuum.
The same fluctuations can be observed in the matrix elements $V_{nj}$ studied in Sec.~\ref{sec:perturbative}. 

For completeness, we note that some states in the $(N_{1\cup 2}, N_{1^\prime})=(1,2)$ sector (in blue), with quasienergy around $\epsilon \approx 2.2$, also exhibit size-independent ${\mathcal I}_{k=0}$, even though they are not 3-particle bound states in the same sense as above.
In this sector, two of the particles are completely localized on the extra sites and then serve as impurity potentials for the third particle that resides on the chain. 
This particle can be either in an extended state on the chain giving ${\mathcal I}_{k=0} \sim \Luc^{-1}$ (these states show collapse in the ${\mathcal I}_{k=0} \cdot \Luc$ panel in Fig.~\ref{fig:IPR_resolved}), or it can be localized on one of the impurities giving size-independent ${\mathcal I}_{k=0}$; it is the latter states, whose details also depend on the relative location of the two impurities, that show up near $\epsilon \approx 2.2$.

\subsection{Continuum states}
In Fig.~\ref{fig:IPR_resolved} we observe three types of 2+1 continuum states that show collapse in the ${\mathcal I}_{k=0} \cdot \Luc$ panel.
First, we find a set of states in the $(N_{1\cup 2}, N_{1^\prime}) = (3, 0)$ sector (red) with quasienergies from $0.8$ to $-2.6+2\pi$.
Another set of states is in the $(N_{1\cup 2}, N_{1^\prime}) = (2, 1)$ sector (green) with quasienergies from $1.9$ to $-3+2\pi$: these can be of two subtypes, one where the 2-particle bound state has one particle localized on an extra site (the third particle roams on the chain), and the other where the 2-particle bound state roams on the chain (the third particle is localized on an extra site); we do not try to distinguish these here.
Finally, most of the states in the $(N_{1\cup 2}, N_{1^\prime}) = (1, 2)$ sector (blue) with quasienergies from $-1$ to $1$ show collapse in the ${\mathcal I}_{k=0} \cdot \Luc$ panel: more precisely, these states are not 2+1 states but instead have two particles completely localized on the extra sites and the third particle extended around the chain, which we have already mentioned in the previous subsection.

Turning to the states that show collapse in the ${\mathcal I}_{k=0} \cdot \Luc^2$ panel, we see the 1+1+1 continuum, that spans the full quasienergy range in the $(N_{1\cup 2}, N_{1^\prime}) = (3,0)$ sector (red), and a range of quasienergies from $-2.1$ to $2.1$ in the $(N_{1\cup 2}, N_{1^\prime}) = (2, 1)$ sector (green) (strictily speaking, the latter are not 1+1+1 continuum but are from flat bands obtained by constructing momentum eigenstates from degenerate states where one of the particles is completely localized on an extra site while the other two are in extended states on the chain). 

Note that in the quasienergy range from $1.2$ to $2$ we find many eigenstates in the $(N_{1\cup 2}, N_{1^\prime}) = (2, 1)$ sector with values of ${\mathcal I}_{k=0}$ that are intermediate between the 2+1 and 1+1+1 continuum: since this sector is not integrable, the states in the 2+1 can decay in the 1+1+1 continuum (if they have the same quasienergy), so the scattering states will be a mixture of 2+1 and 1+1+1 states.
We have not tried to understand these states in any detail since they are far from the 3-particle bound states of main interest to us.

\end{document}